\documentclass[preprint,3p,12pt]{elsarticle}
\usepackage{mathrsfs}
\usepackage{amsmath}
\usepackage{stmaryrd}
\usepackage{bbding}
\usepackage{dcolumn}
\usepackage{graphicx}
\usepackage{amsfonts}
\usepackage{amssymb}
\usepackage{psfrag}
\usepackage{wrapfig}
\usepackage{subfigure}
\usepackage{bm}
\usepackage{epsf}
\usepackage{epsfig}
\usepackage{setspace}
\usepackage{graphicx}
\usepackage{epstopdf}
\usepackage{psfrag}
\usepackage{subfigure}
\usepackage{color}
\usepackage{comment}

\usepackage{float}
\graphicspath{{./image/}}
\usepackage{epstopdf}
\numberwithin{figure}{section}
\numberwithin{equation}{section}

\begin{document}
\title{High-order Unified Gas-kinetic Scheme}

\author[HKUST1]{Gyuha Lim}
\ead{glim@connect.ust.hk}

\author[HKUST2]{Yajun Zhu}
\ead{mazhuyajun@ust.hk}

\author[HKUST1,HKUST2]{Kun Xu\corref{cor1}}
\ead{makxu@ust.hk}

\address[HKUST1]{Department of Mechanical and Aerospace Engineering, The Hong Kong University of Science and Technology, Clear Water Bay, Kowloon, Hong Kong, China}
\address[HKUST2]{Department of Mathematics, The Hong Kong University of Science and Technology, Clear Water Bay, Kowloon, Hong Kong, China}
\cortext[cor1]{Corresponding author}

\begin{abstract}
In this paper, we present a high-order unified gas-kinetic scheme (UGKS) using the weighted essentially non-oscillatory with adaptive-order (WENO-AO) method for spatial reconstruction and the two-stage fourth-order scheme for time evolution.
Since the UGKS updates both the macroscopic flow variables and microscopic distribution function, and provides an adaptive flux function by combining the equilibrium and non-equilibrium parts, it is possible to take separate treatment of the equilibrium and non-equilibrium calculation in the UGKS for the development of high-order scheme. Considering the fact that high-order techniques are commonly required for continuum flow with complex structures, and the rarefied flow structure are relatively simple and smooth in the physical space, we apply the high-order techniques in the equilibrium part of the UGKS for the capturing of macroscopic flow evolution, and retain the calculation of distribution function as a second-order method, so that a balance of computational cost and numerical accuracy could be well achieved. The high-order UGKS has been validated by several numerical test cases, including sine-wave accuracy test, sod-shock tube, Couette, oscillating Couette, lid-driven cavity and oscillating cavity flow. It is shown that the current method preserves the multiscale property of the original UGKS and obtains more accurate solutions in several cases. 
\end{abstract}


\maketitle

\section{Introduction}	
The gas-kinetic scheme (GKS) is a hydrodynamic flow solver based on the kinetimc model equations \cite{BGK, gks1993, gks1994, gks2001}. With the Chapman-Enskog expansion \cite{CE-expansion}, the GKS can recover the Navier-Stokes (NS) solutions, and it combines the upwind and central difference schemes automatically with multidimensionality. However, the use of Chapman-Enskog expansion constrains the application of the GKS only in the continuum flow regime. In order to extend the scheme for non-equilibrium flow, unified gas kinetic scheme (UGKS) has been developed \cite{ugks}. The UGKS is a multiscale flow solver based on the direct modeling of flow physics on the numerical mesh size and time step scale with a discretized particle velocity space, and it utilizes integral solution of the BGK-type model, such as Shakhov equation, for gas evolution and flux evaluation at a cell interface. With the variation of the ratio between the local particle mean collision time and the discrete time step, the multiscale property of the UGKS is achieved with the fully recovering of different flow regimes from free particle transport to the hydrodynamic scale. In comparison with the direct simulation Monte Carlo (DSMC) method \cite{DSMC-Bird}, which is the most prevailing particle method for rarefied flow simulation, the UGKS has advantages in the near continuum regime in terms of high efficiency and accuracy without statistical noises. This property makes the UGKS to be suitable for low speed slip and transition regime problem such microflow and  micro-electro-mechanical system (MEMS) applications \cite{ugks-microflow}. 

In recent years, many high-order methods have been developed in the computational fluid dynamics (CFD) and more accurate solutions are expected to be obtained than the first- and second-order solutions. For the finite volume scheme, the essentially non-oscillatory (ENO) and weighted essentially non-oscillatory (WENO) have been developed \cite{eno-Harten1986, weno} and there are diversely modified versions of WENO, including WENO-JS \cite{weno-js}, WENO-Z \cite{wenoz} and WENO with adaptive-order (WENO-AO) \cite{wenoao-Balsara2016efficient}. 
High-order GKS (HGKS) have also been developed by incorporating these WENO methods for spatial reconstruction \cite{weno-gks, wenoao-gks-Ji2019-performance-enhancement}. Furthermore, the existence of time derivative term in the flux function enables the GKS to provide a high-order time evolution solution with less stages. For instance, two-stage fourth-order temporal discretization method has been applied in the HGKS \cite{s2o4-Li2016, s2o4-Pan2016}, with fewer stages in one step, it achieves better computational efficiency than the Runge-Kutta (RK) method.

While the high-order method successfully implemented for continuum regime, the high-order method for rarefied flow regime have also been investigated in recent years. The original DSMC has first-order accuracy due to the decoupling treatment of convection and collision terms. The higher-order DSMC method was constructed by improving the temporal accuracy of the collision term \cite{DSMC-highorder}. However, statistical noise from particle method is still not resolved to get accurate solution. The discrete unified gas kinetic scheme (DUGKS) \cite{dugks2013, dugks2015} implementing the discrete form of the analytical solution with the coupling of the particles' transport and collision gets high-order solution with third-order accuracy for low speed isothermal rarefied flow simulation by employing two-stage method \cite{dugks-highorder}. The two-stage method is also applied in the UGKS to achieve a third-order multistage UGKS for both microscopic and macroscopic variables \cite{ugks-thirdorder-inproceeding}. From this study, it seems that high-order method for the updating of the distribution function at discrete velocity point is not necessary due to the huge increment of computational cost for its high-order reconstruction with slight improvement in resolving simple flow structures in rarefied regimes. In addition, the boundary induced discontinuities of distribution function at different discrete velocities will easily spread into the inner domain at different locations for rarefied flow simulations. 

However, for near continuum flow at relatively small Knudsen numbers, both of the particles' free transport and collision play important roles. The complex flow structure in the near continuum regime requires high-order scheme to follow its evolution with high resolution. Since the UGKS updates both microscopic and macroscopic flow variables, and couples particles' free transport and collision in flux function by a combination of equilibrium and non-equilibrium parts, it is possible to implement high-order reconstruction on the equilibrium part of the flow variables while keeping the second-order reconstruction for microscopic distribution function. By this way, the capabilities of resolving complex flow structures in continuum regimes, and capturing non-equilibrium physics in rarefied regimes can be both achieved with a slight increment of computational cost.  It is expected to enhance the accuracy in near continuum regime and show advantages for microflows and MEMS applications. 

The present work is to introduce the implementation of WENO-AO in the equilibrium part of the UGKS. The spatial reconstruction for macroscopic variables will be conducted with WENO5-AO, and van Leer flux limiter is used for microscopic variables. Two-stage fourth-order method is applied for temporal discretization. In Section 2, the UGKS, the WENO-AO reconstruction, the two-stage fourth-order temporal discretization method, and WENO-AO implemented UGKS are introduced. Section 3 presents the numerical simulation results of test cases including accuracy test,  1D Riemann problem, Couette flow and cavity flows by UGKS and WENO-AO implemented UGKS. Conclusions will be drawn in the last section.

\section{Numerical method}	
In this section, a detailed description for unified gas-kinetic scheme and high-order reconstruction with WENO5-AO and two-stage fourth-order method is introduced. 

\subsection{Unified gas kinetic scheme}
UGKS is based on the BGK-type model. For monatomic gas, Shakhov equation is commonly used and two-dimensional Shakhov equation can be written as following 
\begin{equation}\label{BGK}
	f_t + u f_x + v f_y = \frac{f^{+}-f}{\tau}
\end{equation}
where $f$ is the initial gas distribution function and $f^{+}$ is the heat flux modified equilibrium state from initial distribution function $f$ with Shakhov model. Shakhov model is defined as following
\begin{equation}\label{Shakhov_model}
	f^{+}=g\left[1+(1-\mathrm{Pr})\mathbf{c} \cdot \mathbf{q} \left(\dfrac{c^2}{RT}-5\right)/(5pRT)\right] = g+g^{+}
\end{equation}
with random velocity $\mathbf{c}=\mathbf{u} - \mathbf{U}$ and the heat flux $\mathbf{q}$. Shakhov model use Hermite polynomial in the equilibrium state to adjust heat flux to achieve arbitrary Prandtl number. For two-dimensional case, the gas distribution function is a function of space $(x,y)$, time $t$ and particle velocity $(u,v,w)$ in $x$-, $y$- and $z$- directions. The particle collision time $\tau$ is related to the viscosity by
\begin{equation}
\begin{aligned}\label{collsion_time}
	\tau = \dfrac{\mu}{p}
\end{aligned}
\end{equation}\label{f_Maxwell}
where $\mu$ is dynamic viscosity. To neglect molecular rotation and vibration, monatomic gas is considered in this paper. Thus, the equilibrium distribution function of monatmoic gas in 2D case can be expressed as following
\begin{equation}
	g=\rho (\frac{\lambda}{\pi})^{\frac{3}{2}}e^{-\lambda((u-U)^2+(v-V)^2)+w^2}
\end{equation}
where $\rho$ is the density, (U,V) is the macroscopic velocity in $x$ and $y$ directions, $\lambda$ is thermodynamic property which is defined as $\lambda=m/2kT=1/2RT$, $m$ is the molecular mass, $k$ is the Boltzmann constant, $R$ is specific gas constant and $T$ is the temperature. The macroscopic properties (i.e., density $\rho$, momentum ($\rho U$, $\rho V$) and energy density $\rho E$) are related to microscopic gas distribution function as following 
\begin{equation}\label{f_moments}
\left[
\begin{matrix}
	\rho \\
	\rho U \\
	\rho V \\
	\rho E \\
\end{matrix}
\right]
=\int\psi_afd\Xi, \alpha=1,2,3,4,
\end{equation}
where $\psi_a$ is the component of the vector moments 
\begin{equation}\label{vector_components}
\boldsymbol{\psi}=(\psi_1,\psi_2,\psi_3,\psi_4)^T=(1, u, v, \frac{1}{2}(u^2+v^2+w^2))^T
\end{equation}
and $d\Xi$ = $dudvdw$ is the volume element in the velocity space. Due to conservation of mass, momentum and energy during collisions, $f$ and $g$ satisfy the compatibility condition, 
$\int (g-f)\psi_\alpha d\Xi=0, \alpha=1,2,3,4,$
at any point in space and time. 
\\

The unified gas-kinetic scheme is based on the finite volume method with discrete physical space and velocity space. The temporal discretization is also performed by $\delta t$ with CFL condition. The averaged gas distribution in a physical domain $\Omega_{i,j}$ at time $t^n$ in the velocity space $\Omega_{k, l}$, i.e., $dudv$ around the velocity point $(u_k, v_l)$, can be written as 
\begin{equation}\label{discretized_f}
	f(x_i,y_j,t^n,u_k,v_l)=f^{n}_{i,j,k,l}=\frac{1}{\Delta x\Delta y\Delta u\Delta v}\iint_{\Omega_{i,j}}\int_{\Omega_{k, l}}f(x,y,t^n,u,v,w)dxdyd\Xi.
\end{equation}

In the framework of finite volume method, the evolution of the gas distribution function  can be written as
\begin{equation} \label{f_evolution}
	f^{n+1}_{i,j} = f^n_{i,j} + \frac{1}{\Omega_{i,j}} \int^{t^{n+1}}_{t^{n}} \sum_{m=1}^{m=N} \hat{u}_m \hat{f_m}(t) \Delta S_m dt + \frac{1}{\Omega_{i,j}} \int^{t^{n+1}}_{t^{n}} \iint_{\Omega_{i,j}} \frac{f^{+}-f}{\tau} d\Omega dt,
\end{equation}
where $N$ is the total number of interfaces of a control volume, $u_m$ is the particle velocity normal to the cell interface and $\Delta S_m$ is the $m$-th interface length. 

Taking conservative moments $\psi_\alpha$ on Eq.~\eqref{f_evolution}, due to the conservation laws of mass, momentum and energy during particle collision process, the update of conservative variables is described as following
\begin{equation}\label{W_update}
	W^{n+1}_{i,j}=W^n_{i,j}+\dfrac{1}{\Omega_{i,j}}\int^{t^{n+1}}_{t^n}\sum_{m=1}^{m=N }\Delta \mathbf{S}_m\cdot\Delta \mathbf{F}_m(t)dt,
\end{equation}
where $W$ is the cell averaged conservative variables which are density, momentum and energy densities inside each control volume and $\mathbf{F}$ is the macroscopic flux across the cell interface for each cell. The macroscopic flux is computed with the local solution of the kinetic equation. 
\\

UGKS applies time-dependent gas distribution function at the cell interface to compute microscopic and macroscopic fluxes. The distribution function at the cell interface with the $x$-direction as normal direction can be written as following
\begin{equation}\label{f_interface}
\begin{aligned}
	\hat{f}_{i+1/2,k,l}&= f(x_{i+1/2},t,u_k,v_l,w)\\
	=&\frac{1}{\tau}\int^{t^{n+1}}_{t^n}f^+(x',t',u_k,v_l,w)e^{-(t-t')/\tau}dt'\\
	+&e^{-(t-t^n)/\tau}f^n_{0,k,l}(x_{i+1/2}-u_k(t-t^n),t^n,u_k,v_l,w),
\end{aligned}
\end{equation}
where $x'=x_{i+1/2}-u_k(t-t')$ is the particle trajectory, $f^n_{0,k,l}$ is the initial gas distribution function of $f$ at time $t=t^n$ around the cell interface $x_{i+1/2}$ at particle velocity $(u_k, v_l)$ and $f^+=g+g^+$ is Shakhov part which will be evaluated separately. By utilizing the above integral equation, UGKS enables to handle flow physics in different scales from free transport mechanism with initial term $f_0$ to the hydrodynamic scale with the integration of the equilibrium state which represents the particle collision effects leading to Maxwellian. The flow behavior is determined by the ratio of time step and local particle collision time. 

The initial distribution function at the cell interface is evaluated with left and right cell of the interface by
\begin{equation}\label{f_initial}
	f_0(x,t^n,u_k,v_l,w)=f_{0,k,l}(x,0)=
	\begin{cases}
		f^L_{i+1/2,k,l}+\cfrac{\partial f_{i,k,l}}{\partial x}x, & x\leq0, \\
		f^R_{i+1/2,k,l}+\cfrac{\partial f_{i+1,k,l}}{\partial x}x, & x>0,
	\end{cases}
\end{equation}
where van Leer nonlinear limiter is used to obtain $f^L_{i+1/2,k,l}$, $f^R_{i+1/2,k,l}$ and the corresponding slopes. 

The one-to-one correspondence between an equilibrium state and macroscopic flow variable enables to determine an equilibrium state gas distribution function $g$ and macroscopic variable $W$ at the interface. For an equilibrium state $g$ around the cell interface ($x_{i+1/2}=0$, $t=0$), it can be expanded with two slopes,
\begin{equation}\label{f_equilibrium}
	g=g_0[1+(1-\mathrm{H}(x))\bar{a}^Lx+\mathrm{H}[x]\bar{a}^Rx+\bar{A}t],
\end{equation}
where $g_0$ is Maxwellian distribution function at $x=0$, $\bar{a}^L$, $\bar{a}^R$ and $\bar{A}$ are derivative terms of a Maxwellian distribution function in space and time, $\mathrm{H}[x]$ is the Heaviside function  defined as
\begin{equation}\label{Heaviside}
\begin{aligned}
	\mathrm{H}[x]= \left\{ \begin{array}{lcl} 
		0, & \mbox{for}& x<0, \\
		1, & \mbox{for}& x\geq 0,
	\end{array}\right.
\end{aligned}
\end{equation}

The following relation is obtained from the compatibility condition of the BGK model. The conservation constraints at $(x=x_{i+1/2}, t=0)$ provides 
\begin{equation}\label{conservation_constraints}
	W_0=\int g_0\boldsymbol{\psi}d\Xi
	=\sum(f^L_{i+1/2,j,k}\mathrm{H}[u_k]+f^R_{i+1/2,k,l}(1-\mathrm{H}[u_k]))\boldsymbol{\psi},
\end{equation}
where $W_0=(\rho_0,\rho_0U_0,\rho_0V_0,\rho_0E_0)^T$ are the conservative variables, and this moments can be computed explicitly by using initial distribution function at the cell interface. 

The derivative parts of equilibrium, $\bar{a}^L$ and $\bar{a}^R$ can be computed by matrix calculation as following
\begin{equation}\label{spatial_der_right}
\begin{aligned}
	\dfrac{\bar{W}_{i+1}(x_{i+1})-W_0}{\rho_0\Delta x^+} 
	=
	\dfrac{1}{\rho_0}\int\bar{a}^Rg_0\psi d\Xi
	=
	\bar{M}^0_{\alpha\beta}
	\left[	
	\begin{matrix}
		\bar{a}_1^R \\
		\bar{a}_2^R  \\
		\bar{a}_3^R  \\
		\bar{a}_4^R  \\
	\end{matrix}
	\right]	
	=
	\bar{M}^0_{\alpha\beta}\bar{a}^R_\beta,
\end{aligned}
\end{equation}

\begin{equation}\label{spatial_der_left}
\begin{aligned}
	\dfrac{W_0-\bar{W}_{i}(x_{i})}{\rho_0\Delta x^-} 
	=
	\dfrac{1}{\rho_0}\int\bar{a}^Lg_0\psi d\Xi
	=
	\bar{M}^0_{\alpha\beta}
	\left[	
	\begin{matrix}
		\bar{a}_1^L \\
		\bar{a}_2^L  \\
		\bar{a}_3^L  \\
		\bar{a}_4^L  \\
	\end{matrix}
	\right]	
	=
	\bar{M}^0_{\alpha\beta}\bar{a}^L_\beta,
\end{aligned}
\end{equation}
where the matrix is $\bar{M}^0_{\alpha\beta}=\int g_0\psi_\alpha\psi_\beta d\Xi/\rho_0$, $\Delta x^+=x_{i+1}-x_{i+1/2}$ and $\Delta x^-=x_{i+1/2}-x_{i}$ are the distances from the cell center to the cell interface. Then, the time evolution derivative part $\bar{A}$ can be evaluated with following relation
\begin{equation}
\begin{aligned}\label{time_evolution}
	\dfrac{d}{dt}\int(g-\hat{f})\boldsymbol{\psi}d\Xi=0,
\end{aligned}
\end{equation}
at $(x=0,t=0)$ and get 
\begin{equation}\label{temporal_der}
	\begin{aligned}
		\bar{M}^0_{\alpha\beta}\bar{A}_\beta, 
		&= \dfrac{1}{\rho_0}(\partial\rho/\partial t,\partial(\rho U)/\partial t, \partial(\rho V)/\partial t, \partial(\rho E)/\partial t)^T \\	
		&=\dfrac{1}{\rho_0}\int[u(\bar{a}^L\mathrm{H}[u]+\bar{a}^R(1-\mathrm{H}[u]))g_0]\boldsymbol{\psi}d\Xi.\\
	\end{aligned}
\end{equation}

By substituting Eq.~\eqref{f_initial} and Eq.~\eqref{f_equilibrium} into Eq.~\eqref{f_interface} and taking $(u=u_k, v=v_l)$ into $g_0$, $\bar{a}^L$,$\bar{a}^R$ and $\bar{A}$, the gas distribution function $\hat{f}(x_{i+1/2},t,u,v,w)$ at the discretized particle velocity $(u_k,v_l)$ is defined as following
\begin{equation}\label{f_final}
\begin{aligned}
	\hat{f}(x_{i+1/2},t,u,v,w)=&(1-e^{-t/\tau})(g_0+g^+)\\
	+&((t+\tau)e^{-t/\tau}-\tau)(\overline{a}^L\mathrm{H}[u]+\overline{a}^R(1-\mathrm{H}[u]))u_kg_0\\	
	+&\tau(t/\tau-1+e^{-t/\tau})\bar{A}g_0\\
	+&e^{-t/\tau}\left( (f^L_{i+1/2,k}-u_kt\sigma_{i,k})\mathrm{H}[u_k]
	+(f^R_{i+1/2,k}-u_kt\sigma_{i+1,k})(1-\mathrm{H}[u_k]) \right)\\
	\triangleq&\tilde{g}_{i+1/2,k,l}+\tilde{f}_{i+1/2,k,l},
\end{aligned}
\end{equation}
where $\tilde{g}_{i+1/2,k,l}$ is terms related to the equilibrium state $g$ and $g^+$, and $\tilde{f}_{i+1/2,k,l}$ is terms related to initial condition $f_0$.

The UGKS updates macroscopic variables with Eq.~\eqref{W_update} and flux $\mathbf{F}$ is computed as 
\begin{equation}\label{macroscopic_flux}
	\mathbf{F}=\int u\psi\hat{f}_{i+1/2,k,l}d\Xi.
\end{equation}

For the particle collision term, the trapezoidal rule is used for UGKS. Thus, UGKS for the update of gas distribution function is 
\begin{equation}\label{f_update}
\begin{aligned}
	f^{n+1}_{i,j,k,l} = (1+\frac{\Delta t}{2 \tau_{i,j}^{n+1}})^{-1} 
	\left[
	f^n_{i,j,k,l} + \frac{1}{\Omega_{i,j}}  \int^{t^{n+1}}_{t^n}  \sum_{m}  \Delta S_m u_m \hat{f}_{m,k,l}  dt  
	+ \frac{\Delta t}{2} \left(\frac{f^{+(n+1)}_{i,j,k,l}}{\tau^{n+1}_{i,j}} + \frac{f^{+(n)}_{i,j,k,l} - f^n_{i,j,k,l}}{\tau^n_{i,j}} \right) \right]
\end{aligned}
\end{equation}
where no iteration is required for the update of the above solution. To save computational cost, the reduced distribution function is introduced. The particle velocity in z-direction can be integrated into internal motion of the particle for two-dimensional cases. Since this paper considers monatomic gas where no internal motion exist, the two reduced distribution function are following
\begin{equation}
\begin{aligned}\label{reduced_distribution}
	h=\int fdw, \quad b=\int w^2fdw,
\end{aligned}
\end{equation}
When internal degree of freedom exists, it can be integrated into the reduced distribution function. Then, Eq.~\eqref{BGK} becomes 
\begin{equation}
\begin{aligned}\label{h_BGK}
	h_t+uh_x+vh_y=\frac{h^{+}-h}{\tau},
\end{aligned}
\end{equation}
\begin{equation}
\begin{aligned}\label{b_BGK}
	b_t+ub_x+vb_y=\frac{b^{+}-b}{\tau},
\end{aligned}
\end{equation}
The macroscopic variables becomes
\begin{equation}\label{macro_and_reduced}
	W=
	\left(
	\begin{matrix}
		\int h d\Xi \\
		\int uh d\Xi \\
		\int vh d\Xi \\
		\int \frac{1}{2}((u^2+v^2)h+b)d\Xi \\
	\end{matrix}
	\right)
\end{equation}
where $d\Xi=dudv$ in two-dimensional cases. 

\subsection{High-order reconstruction}	

\subsubsection{WENO-AO reconstruction}
The fifth-order WENO-AO reconstruction proposed by Balsara \cite{wenoao-Balsara2016efficient} on a uniform rectangular mesh is presented in this section. The WENO5-AO formulation is based on one-dimensional cases in this paper. fifth-order spatial accuracy is selected to pair with fourth-order temporal accuracy. 

Assume that $\overline{Q}$ are the cell-averaged variables, and $Q$ are the reconstructed variables and conservative variables are used for the reconstruction in this paper. Three sub-stencils are used to achieve fifth-order spatial accuracy of the reconstructed value. This paper will take the left interface value $Q_{i+1/2}^l$ of the cell interface $x_{i+1/2}$ as the example and explain. The sub-stencils are chosen as following
\begin{equation}\label{substencil}
	S_0=\{I_{i-2},I_{i-1}, I_i\}, ~~S_1=\{I_{i-1},I_i, I_{i+1}\}, ~~S_2=\{ I_i,
	I_{i+1},I_{i+2}\}.
\end{equation}
For each sub-stencil $S_k$, a unique quadratic  polynomial $p^{r3}_{k}(x)$ are evaluated by $\overline{Q}$, and they are constructed by
\begin{equation}\label{polynomial_substencil}
	\frac{1}{\Delta x}\int_{I_{i-j-k-1}}p^{r3}_{k}(x)dx=\overline{Q}_{i-j-k-1},~j=-1,0,1,
\end{equation}
Each $p^{r3}_{k}(x)$ can achieve a third-order spatial accuracy in smooth flow region. By taking $x_{i+1/2}$ into $p^{r3}_{k}(x)$, the reconstructed point-wise values are evaluated as following
\begin{equation}\label{pointwise_substencil}
\begin{aligned}
	p_{0}^{r3}(x_{i+1/2})&=\frac{1}{3}\overline{Q}_{i-2}-\frac{7}{6}\overline{Q}_{i-1}+\frac{11}{6}\overline{Q}_{i},\\	
	p_{1}^{r3}(x_{i+1/2})&=-\frac{1}{6}\overline{Q}_{i-1}+\frac{5}{6}\overline{Q}_i+\frac{1}{3}\overline{Q}_{i+1},\\	
	p_{2}^{r3}(x_{i+1/2})&=\frac{1}{3}\overline{Q}_{i}+\frac{5}{6}\overline{Q}_{i+1}-\frac{1}{6}\overline{Q}_{i+2}.	
	\end{aligned}
\end{equation}
A large stencil, $\mathbb{S}_3 = \{S_0, S_1, S_2\}$, which includes all three sub-stencils, will also have a unique fifth-order polynomial $p_3^{r5}(x)$ and the polynomial is constructed as following
\begin{equation}\label{large_stencil}
	\frac{1}{\Delta x}\int_{I_{i+j}}p_3^{r5}(x)dx=\overline{Q}_{i+j}, ~j=-2,-1, 0,1,2.
\end{equation}
With the above formulation, the corresponding point-wise value at the cell interface $x_{i+1/2}$ is evaluated as following
\begin{equation}\label{r5_pointwise}
	p_{3}^{r5}(x_{i+1/2})=\frac{1}{60}(47\overline{Q}_{i}-13\overline{Q}_{i-1}+2\overline{Q}_{i-2}+27\overline{Q}_{i+1}-3\overline{Q}_{i+2}).
\end{equation}
The weight $d_k, k=0,1,2,$ for each sub-stencil is evaluated as following
\begin{equation}\label{r5_evaluation}
	p_{3}^{r5}(x_{i+1/2})=\sum_{k=0}^{2}d_{k}p_{k}^{r3}(x_{i+1/2}),
\end{equation}
where $d_k$ are unique, and $\displaystyle d_{0}=\frac{1}{10},d_{1}=\frac{3}{5}, d_{2}=\frac{3}{10}$.
\\

After obtaining $p_{3}^{r5}(x)$ and $p_{k}^{r3}(x)$, $k=0,1,2$, the fifth-order polynomial for whole stencil $p_3^{r5}(x)$ is written as following
\begin{equation}\label{rewrite-r5}
	p_{3}^{r5}(x)=\gamma_3(\frac{1}{\gamma_3}p_3^{r5}(x)-\sum_0^2 \frac{\gamma_k}{\gamma_3}p_k^{r3}(x))+\sum_0^2 {\gamma_k}p_k^{r 3}(x),
\end{equation}
where $\gamma_{k}, k=0,1,2,3$ are linear weights, and its value is evaluated by Balsara et al. \cite{wenoao-Balsara2016efficient},
\begin{equation}\label{linear_weights}
	\gamma_3=\gamma_{Hi},~~ \gamma_0=\gamma_2=(1-\gamma_{Hi})(1-\gamma_{Lo})/2, ~~\gamma_1=(1-\gamma_{Hi})\gamma_{Lo},
\end{equation}
where $\gamma_{Hi} \in [0.85,0.95]$ and $\gamma_{L_o} \in [0.85,0.95]$. The sum of linear weights satisfies $\sum_0^3 \gamma_{k} = 1$ and $\gamma_{k}>0, k=0,1,2,3$. If there is no specification about linear weights, $\gamma_{Hi} =0.85 $ and $\gamma_{l_o} =0.85 $ are adopted. 
\\

For the nonlinear weights, the WENO-Z type \cite{wenoz} is selected and they are evaluated as following
\begin{equation}\label{nonlinear_weights}
	\omega_k=\gamma_k(1+ \frac{\tau_s^2}{(\beta_{k}+\epsilon)^2} ),
\end{equation}
where $\tau_s$ is the global smooth indicator, and it is defined as 
\begin{equation}\label{smooth_indicator}
	\tau=\frac{1}{3}(|\beta_3^{r5}-\beta_0^{r3}|+|\beta_3^{r5}-\beta_1^{r3}|+|\beta_3^{r5}-\beta_2^{r3}|) = O(\Delta h^4).
\end{equation}
where $\beta_{k}=\beta_k^{r3}$, $k=0,1,2$, is the smooth indicator of sub-stencil $S_k$, and $\beta_3=\beta_3^{r5}$ is the smooth indicator of the whole stencil $\mathbb{S}_3$. Balsara et al.\cite{wenoao-Balsara2016efficient} provides the explicit formula for the $\beta_{k}$. $\epsilon$ is a positive small number to avoid zero for denominator, and $\epsilon = 10^{-6}$ is selected in whole paper. 
Then, normalization is performed for the weights $\overline{\omega}_k$ as following
\begin{equation}\label{nonlinaer_weights_normalization}
	\overline{\omega}_k=\frac{\omega_k}{\sum_0^3 \omega_q}.
\end{equation}
The final form of the reconstructed polynomial is written as following
\begin{equation}\label{weno-ao-re-polynomial}
	P^{AO(5,3)}(x)=\overline{\omega}_3(\frac{1}{\gamma_3}p_3^{r5}(x)-\sum_0^2 \frac{\gamma_k}{\gamma_3}p_k^{r3}(x))+\sum_0^2 {\overline{\omega}_k}p_k^{r 3}(x).
\end{equation}
The reconstructed left interface value $Q_{i+1/2}^l$ of the cell interface $x_{i+1/2}$ and the corresponding derivative is written as following
\begin{equation}\label{left_reconstruction}
	Q^l_{i+1/2}=P^{AO(5,3)}(x_{i+1/2}), ~~(Q^l_{x})_{i+1/2}=P_x^{AO(5,3)}(x_{i+1/2}).
\end{equation}
With the similar approach, the right interface value $Q^r_{i-1/2}$ of the cell interface $x_{i-1/2}$ and its derivative is also evaluated as following
\begin{equation}\label{right_reconstruction}
	Q^r_{i-1/2}=P^{AO(5,3)}(x_{i-1/2}), ~~ (Q^r_{x})_{i-1/2}=P_x^{AO(5,3)}(x_{i-1/2}).
\end{equation}

The reconstructed value and its normal derivative can be obtained by the above procedure. While the GKS has the multi-dimensional property, not only for the normal derivative $\left(Q_x\right)$ but also $\left(Q_y, Q_z\right)$ is needed for two-dimensional and three-dimensional cases. To preserve multi-dimensional property of GKS and UGKS, the multi-dimensional reconstruction is performed for two-dimensional and three-dimensional cases. The details of multi-dimensional WENO-AO reconstruction procedure may refer to \cite{wenoao-gks-Ji2019-performance-enhancement}.

\subsubsection{Two-stage fourth-order temporal discretization}
The two-stage fourth-order temporal discretization is usually applied to high-order GKS\cite{s2o4-Pan2016}. To pair with high-order spatial discretization, the two-stage fourth-order temporal discretization is applied to WENO-AO implemented UGKS. The second-order flux function in GKS enables to achieve fourth-order temporal accuracy within two steps. For the time-dependent equation,
\begin{equation}\label{pde}
	\frac{\partial W}{\partial t}=\mathcal {L}(W),
\end{equation}
with the initial condition at $t_n$,
\begin{equation}\label{pde2}
	W(t=t_n)=W^n,
\end{equation}
where $\mathcal {L}$ is an operator for spatial derivative terms of flux. The time derivatives can be obtained by the Cauchy-Kovalevskaya method, 
\begin{equation}\label{CK_method}
	\frac{\partial W^n}{\partial t}=\mathcal{L}(W^n), \quad \frac{\partial }{\partial t}\mathcal
	{L}(W^n)=\frac{\partial }{\partial W}\mathcal
	{L}(W^n)\mathcal {L}(W^n).
\end{equation}
An intermediate stage at $t_*=t_n + \Delta t/2$ is required for the two-stage fourth-order method. 
\begin{equation}\label{intermediate_stage}
	W^*=W^n+\frac{1}{2}\Delta t\mathcal{L}(W^n)
	+\frac{1}{8}\Delta t^2\frac{\partial}{\partial t}\mathcal{L}(W^n),
\end{equation} 
The time derivatives for the intermediate state is obtained by
\begin{equation}\label{time_der_intermediate}
	\frac{\partial W^*}{\partial t}=\mathcal{L}(W^*),
	\frac{\partial }{\partial t}\mathcal{L}(W^*)=
	\frac{\partial }{\partial W}\mathcal{L}(W^*)\mathcal {L}(W^*).
\end{equation}
Then, a fourth-order temporal accurate solution for $W(t)$ at $t=t_n +\Delta t$ is updated as following
\begin{equation}\label{update_n+1}
	W^{n+1}=W^n+\Delta t\mathcal{L}(W^n)
	+\frac{1}{6}\Delta t^2\big(\frac{\partial}{\partial t}\mathcal{L}(W^n)
	+2\frac{\partial}{\partial t}\mathcal{L}(W^*)\big).
\end{equation}
The detailed proof can refer to \cite{grp}.
The time-dependent flux is expanded as
\begin{equation}\label{expansion}
	\textbf{F}_{i+1/2,j}(\textbf{W}^n,t)=\textbf{F}_{i+1/2,j}^n+\partial_t \textbf{F}_{i+1/2,j}^n \left(t-t_n\right), t\in\left[t_n,t_n+\Delta t\right].
\end{equation}
To get coefficients of $\textbf{F}_{i+1/2,j}^n$ and $\partial_t\textbf{F}_{i+1/2,j}^n$, the following notation is introduced
\begin{equation}\label{new_notation}
\begin{aligned}
	\mathbb{F}_{i+1/2,j}(W^n,\delta)
	&=\int_{t_n}^{t_n+\delta}\textbf{F}_{i+1/2,j}(W^n,t)dt \\
	&=\sum_{\ell=1}^2\omega_\ell \int_{t_n}^{t_n+\delta}\int u \bm{\psi} f(x_{i+1/2,j_\ell},t,u,v,w)d\Xi dt.
	\end{aligned}
\end{equation}
In the above equation, let $\delta$ as $\Delta t$ and $\Delta t/2$. Then, the equation is written as following
\begin{equation}\label{apply_half}
\begin{aligned}
	\textbf{F}_{i+1/2,j}(W^n,t_n)\Delta t &+\frac{1}{2}\partial_t
	\textbf{F}_{i+1/2,j}(W^n,t_n)\Delta t^2 =\mathbb{F}_{i+1/2,j}(W^n,\Delta t) , \\
	\frac{1}{2}\textbf{F}_{i+1/2,j}(W^n,t_n)\Delta t&+\frac{1}{8}\partial_t
	\textbf{F}_{i+1/2,j}(W^n,t_n)\Delta t^2 =\mathbb{F}_{i+1/2,j}(W^n,\Delta t/2).
	\end{aligned}
\end{equation}
By solving the linear equation above, the coefficient can be computed. 
\begin{equation}\label{get_coefficients}
\begin{aligned}
	\textbf{F}_{i+1/2,j}(W^n,t_n)&=(4\mathbb{F}_{i+1/2,j}(W^n,\Delta t/2)-\mathbb{F}_{i+1/2,j}(W^n,\Delta t))/\Delta t, \\
	\partial_t \textbf{F}_{i+1/2,j}(W^n,t_n)&=4(\mathbb{F}_{i+1/2,j}(W^n,\Delta t)-2\mathbb{F}_{i+1/2,j}(W^n,\Delta t/2))/\Delta
	t^2.
\end{aligned}
\end{equation}
The coefficients for the intermediate state $\textbf{F}_{i+1/2,j}(W^*,t_*)$, $\partial_t \textbf{F}_{i+1/2,j}(W^*,t_*)$ is computed in the same way. Thus, the final flux for the update of intermediate state $W^*_{ij}$ is following
\begin{equation}\label{final_flux_for_interdediate}
	\mathscr{F}_{i+1/2,j}^*=\frac{1}{2}\textbf{F}_{i+1/2,j}(W^n,t_n)+\displaystyle\frac{\Delta	t}{8}\partial_t \textbf{F}_{i+1/2,j}(W^n,t_n),
\end{equation}
Then, the flux for update of next time step $W^{n+1}_{ij}$ is following
\begin{equation}\label{final_flux_for_nextstep}
	\mathscr{F}_{i+1/2,j}^n=\textbf{F}_{i+1/2,j}(W^n,t_n)+\displaystyle\frac{\Delta
		t}{6}\big[\partial_t \textbf{F}_{i+1/2,j}(W^n,t_n)+2\partial_t \textbf{F}_{i+1/2,j} (W^*,t_*)\big].
\end{equation}
More detailed procedure for two-stage fourth-order method can refer to \cite{s2o4-Pan2016}.

\subsection{WENO-AO implemented unified gas kinetic scheme}
In this section, a WENO-AO implemented UGKS will be presented based. The main difference between original second-order UGKS and WENO-AO implemented UGKS is the evaluation of flux terms related to equilibrium state $g_0$. For original second-order UGKS, flux related to macroscopic equilibrium terms is computed from microscopic distribution function according to Eq.~\eqref{conservation_constraints}. Instead of using microscopic variables, the WENO-AO UGKS uses WENO-AO reconstruction with two-step fourth-order method for the macroscopic variables during the flux calculation. Thus, the WENO-AO implemented UGKS have two separate reconstruction procedures. (i.e., van Leer flux limiter for distribution functoin and WENO-AO for macroscopic variables) 

The procedures to evolve the flow field by the one-dimensional WENO-AO implemented UGKS from $t^n$ to $t^{n+1}$ are explained. 
\\

Step 1. Reconstruction of reduced distribution function
\\
Using van Leer limiter with the initial distribution function in each cell, perform spatial interpolation to compute the spatial derivatives of distribution function. Then, get the distribution function at the interface using spatial interpolation. Thus, the distribution function at the interface is from left and right cell depending on the normal particle velocity of the interface.  
\\ 

Step 2. Reconstruction of macroscopic flow variables
\\
Using WENO-AO reconstruction with the macroscopic variables in each cell, calculate the macroscopic variables and its spatial derivatives at the interface. Since WENO-AO reconstruction provides different left and right value ($W^l$ and $W^r$) and its spatial derivatives ($W^l_{x_i}$ and $W^r_{x_i}$) at the interface with different choice of stencils, compatibility condition is used to calculate equilibrium as following
\begin{equation}\label{compatiblity_cond_for_macroscopic}
	W_0 = \int_{u>0}\int \psi_\alpha g^ld\Xi+\int_{u<0}\int \psi_\alpha g^rd\Xi,
\end{equation}
where $g_l$ and $g_r$ are corresponding Maxwellian from $W^l$ and $W^r$. For the spatial derivatives, similar approach is adopted. Compute corresponding slope $\bar{a}^l$ and $\bar{a}^r$ from $W^l_{x_i}$ and $W^r_{x_i}$ using matrix calculation, and apply compatibility condition as following 
\begin{equation}\label{compatiblity_cond_for_der}
	\frac{\partial W_0}{\partial x_i} = \int_{u>0}\int \psi_\alpha \bar{a}^l_ig^ld\Xi+\int_{u<0}\int \psi_\alpha \bar{a}^r_ig^rd\Xi,
\end{equation}
Then, compute temporal derivative of macroscopic variables as following
\begin{equation}\label{compatbility_cond_for_temporal}
	\frac{\partial W_0}{\partial t}= - \rho_0 u_i\frac{\partial W_0}{\partial x_i}
\end{equation}
From temporal derivative, get corresponding slope $\bar{A}$. Thus, all reconstructions for flux calculation are done. 
\\

Step 3. Flux calculation
\\
There is minor change with distribution function at the interface for the WENO-AO implemented UGKS due to combined Maxwellian slope expression. While the distribution function at the interface for original second-order UGKS is given as Eq.~\eqref{f_final}, the distribution function for WENO-AO implemented UGKS is following

\begin{equation}\label{f_final_WENOimplemented}
	\begin{aligned}
		\hat{f}(x_{i+1/2},t,u,v,w)=&(1-e^{-t/\tau})(g_0+g^+)\\
	+&((t+\tau)e^{-t/\tau}-\tau)\overline{a}u_kg_0\\	
	+&\tau(t/\tau-1+e^{-t/\tau})\bar{A}g_0\\
	+&e^{-t/\tau}\left( (f^L_{i+1/2,k}-u_kt\sigma_{i,k})\mathrm{H}[u_k]
	+(f^R_{i+1/2,k}-u_kt\sigma_{i+1,k})(1-\mathrm{H}[u_k]) \right)\\
	\triangleq&\tilde{g}_{i+1/2,k,l}+\tilde{f}_{i+1/2,k,l},
	\end{aligned}
\end{equation} 
Then, compute microscopic and macroscopic flux across the interface with the reconstructed distribution function. 

Furthermore, to apply two-stage fourth-order method, calculate the flux with $\Delta t$ and $\Delta t/2$. Compute intermediate stage $W^*$ by using flux with $\Delta t/2$ and repeat from the reconstruction stage to calculate flux at the intermediate stage. Then, get the final flux with Eq.~\eqref{final_flux_for_nextstep}
\\

Step 4. Update of variables
\\
Update the conservative variables $W^{n+1}$ with the conservation laws Eq.~\eqref{W_update}, and get the corresponding equilibrium state $g^{n+1}$. Then, update the distribution function with Eq.~\eqref{f_update}. 
\\

While the WENO-AO implemented preserves multiscale solving property with discretized particle velocity space, accuracy of the WENO-AO implemented UGKS provides high-order near continuum regime. Especially for low Knudsen number cases, the WENO-AO implemented UGKS could provide accurate solution within less mesh number. 
\\

\section{Numerical results}
This section provides several numerical tests for one-dimensional (1-D) and two-dimensional (2-D) cases. For all test cases, the time step, $\Delta t$ is determined by the CFL condition with CFL number 0.5, and WENO5-AO is used for reconstruction of the conservative variables in the UGKS. 

\subsection{1-D test cases}
\subsubsection{1-D sine wave accuracy test}
The advection of density perbutation is computed to validate the order of numerical scheme. The physical domain is set as $[0,2]$, and the initial condition is given as
\begin{equation}\label{1d_accuracy_initial}
	\rho(x) = 1+0.2\sin(\pi x),\ U(x) = 1, \ p(x) = 1.
\end{equation}
With the periodic boundary conditions at each end, the exact solution for this test case is
\begin{equation}\label{1d_accuracy_analytic}
	\rho(x,t) = 1+0.2\sin(\pi (x-t)), \ U(x,t) = 1,\ p(x,t)=1, 
\end{equation}
In the UGKS, Prandtl number $Pr = 1$ is used to reduce Shakhov equation to BGK equation, and Knudsen number $Kn = 10^{-12}$ is used to ensure the inviscid Euler limit. 
Velocity space is discretized with 201 points with maximum of 10 and minimum of -10. 
A uniform mesh with $N$ points is generated for each calculation. 
The $L^1$, $L^2$, and $L^\infty$ errors and the corresponding orders at $t = 2$ are tabulated in Table~\ref{Table:1D-accuracy-second} and Table~\ref{Table:1D-accuracy-weno}. 

\begin{table}[]
	\centering
	\caption{Accuracy test for the 1-D sine wave propagation by the original second-order UGKS with smooth solver.}
	\label{Table:1D-accuracy-second}
	\begin{tabular}{||c|cc|cc|cc||} \hline
		mesh length	&$L^1$ error  &Order  &$L^2$ error  &Order &$L^\infty$ error  &Order  \\ \hline
		1/10		&5.92807E-02  &  &5.26378E-02  &  &6.51379E-02  &  \\
		1/20		&2.28222E-02  &1.377125   &1.90793E-02  &1.464091   &2.33962E-02  &1.477223   \\ 
		1/40		&7.46302E-03  &1.612606   &6.19637E-03  &1.622513   &9.11200E-03  &1.360435   \\
		1/80		&1.94267E-03  &1.941719   &1.86602E-03  &1.731459   &3.48012E-03  &1.388631   \\
		1/160		&4.77504E-04  &2.024456   &5.49233E-04  &1.764474   &1.30862E-03  &1.411091   \\
		1/320		&1.16179E-04  &2.039164   &1.61136E-04  &1.769140   &5.05430E-04  &1.372463   \\ \hline
	\end{tabular}
\end{table}

\begin{table}[]
	\centering
	\caption{Accuracy test for the 1-D sine wave propagation by the WENO5-AO implemented UGKS with smooth solver. The linear weights of $\gamma_{Hi} = 0.85$, $\gamma_{Lo} = 0.85$ used.}
	\label{Table:1D-accuracy-weno}
	\begin{tabular}{||c|cc|cc|cc||}  \hline
	mesh length	&$L^1$ error  &Order  &$L^2$ error  &Order &$L^\infty$ error  &Order  \\ \hline
1/10		&1.89390E-03  &  &1.52592E-03  &  &1.49810E-03  &  \\
1/20		&6.33295E-05  &4.902339   &4.97763E-05  &4.938077   &5.20243E-05  &4.847805   \\ 
1/40		&1.99965E-06  &4.985058   &1.56766E-06  &4.988774   &1.64425E-06  &4.983684   \\
1/80		&6.25525E-08  &4.998536   &4.90395E-08  &4.998525   &5.15015E-08  &4.996671   \\
1/160		&1.95451E-09  &5.000189   &1.53222E-09  &5.000249   &1.60993E-09  &4.999545   \\
1/320		&6.10576E-11  &5.000492   &4.78699E-11  &5.000361   &5.03320E-11  &4.999378   \\ \hline
	\end{tabular}
\end{table}
The original second-order UGKS and the high-order UGKS with WENO5-AO reconstruction are compared.
From Table~\ref{Table:1D-accuracy-second} and Table~\ref{Table:1D-accuracy-weno}, it can be found that the original second-order UGKS gives maximum order of accuracy 2, while WENO5-AO implemented UGKS provides fifth-order of accuracy as expected. 

\subsubsection{Sod shock tube}
The Sod shock tube test case \cite{Sod1978} with three different Knudsen numbers, Kn = $10, 10^{-3}, 10^{-5}$, is computed. For the computational domain in $x \in [0,1],$ the initial condition is
\begin{equation}\label{sod_initial}
(\rho, U, p) = 
\begin{cases}
	(1,0,1), \quad & \text{for} \quad 0<x<0.5, \\
	(0.125,0,0.1), \quad & \text{for} \quad  0.5\leq x<1,
\end{cases}
\end{equation}
The left and right boundary conditions are set as its initial condition with Maxwellian ghost cells. For comparison, both the second-order UGKS and high-order UGKS adopt the Shakhov model. The hard sphere model is used for the monatomic gas with Pr number $2/3$. The physical space is discretized into 100 cells. and the velocity space is discretized into 200 points in the range of $[-5\sqrt{2k_BT_L/m},5\sqrt{2k_BT_L/m}]$ with trapezoidal rules. The output at $t = 0.15(L\sqrt{m/(2k_BT)})$ is compared. 
\\
\begin{figure}[H]
	\includegraphics[width=0.5\textwidth]{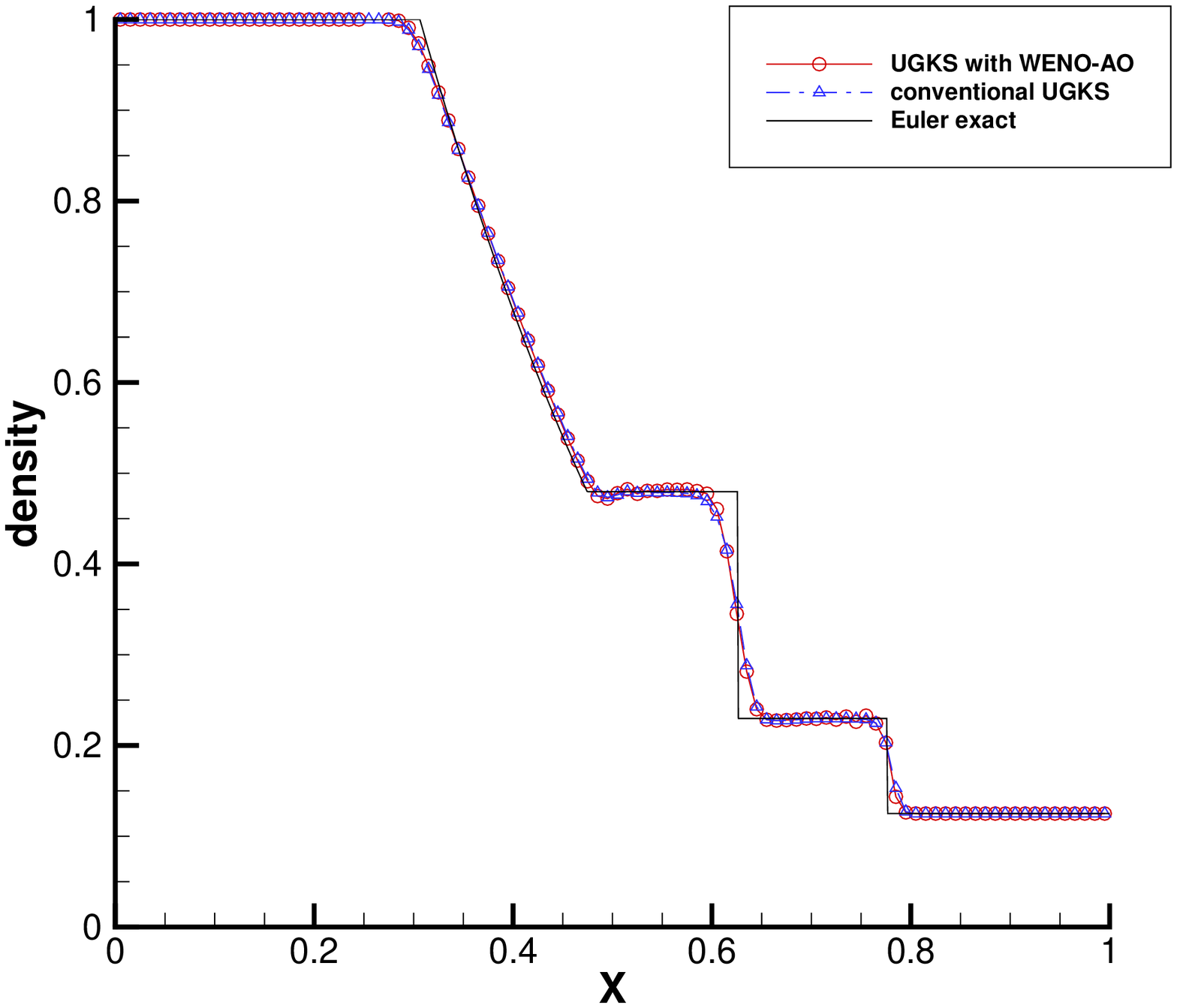}
	\includegraphics[width=0.5\textwidth]{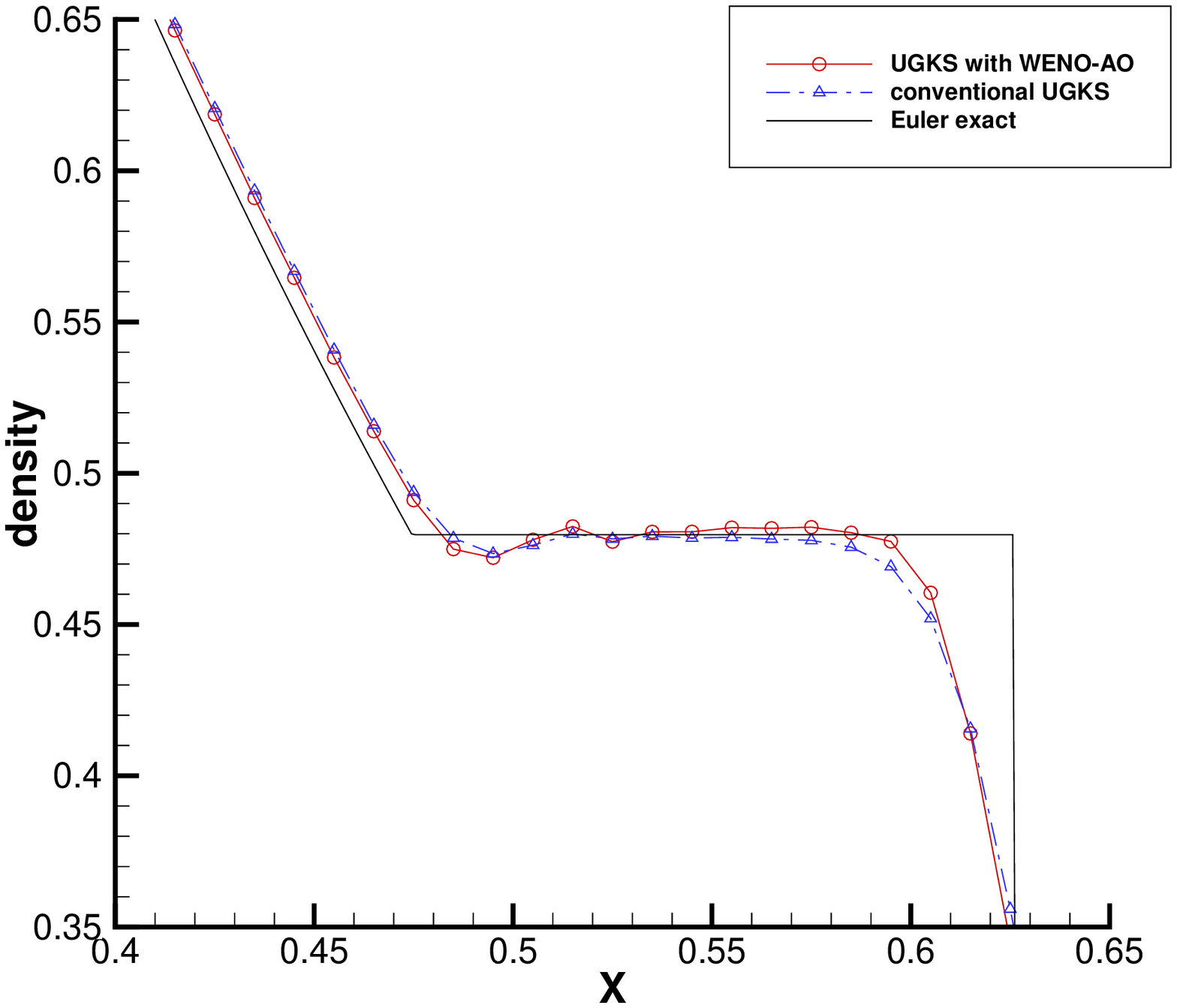}
	\caption{Sod shock tube: the density distribution and local enlargements at Kn = $10^{-5}$} 
	\label{Fig:sod_1e-05}
\end{figure}

\begin{figure}[H]
	\includegraphics[width=0.5\textwidth]{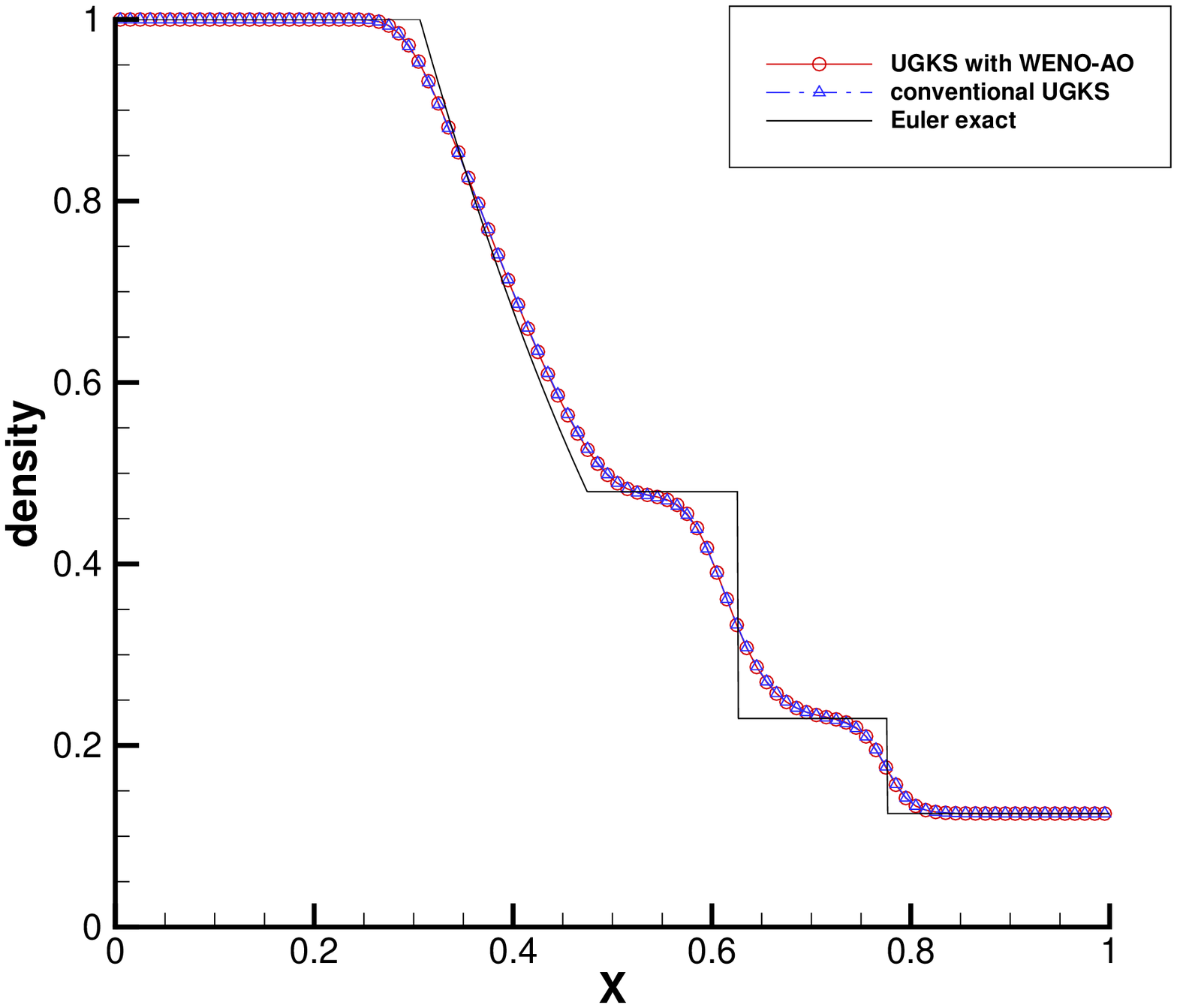}
	\includegraphics[width=0.5\textwidth]{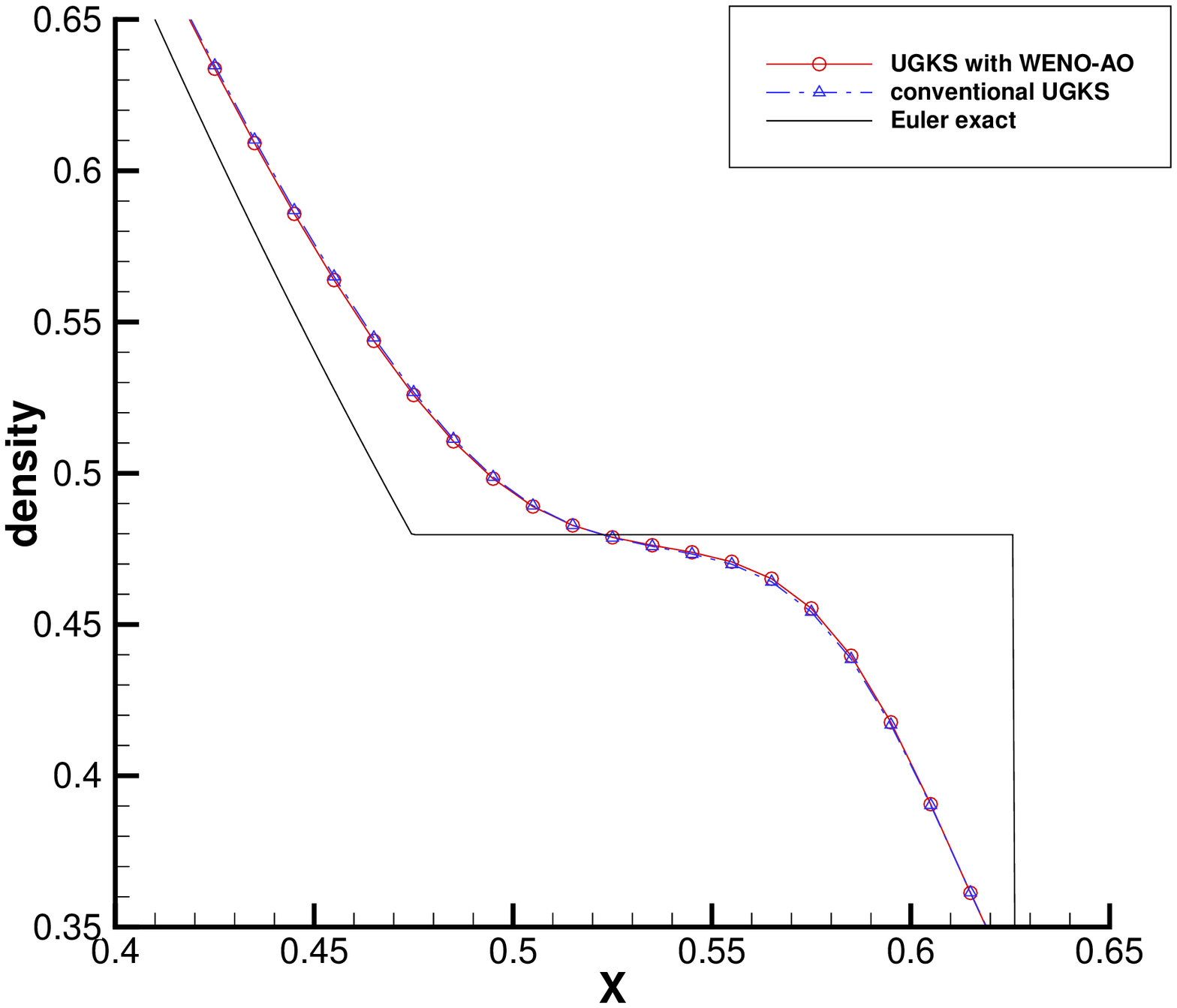}
	\caption{Sod shock tube: the density distribution and local enlargements at Kn = $10^{-3}$} 
	\label{Fig:sod_1e-03}

\end{figure}
\begin{figure}[H]
	\includegraphics[width=0.5\textwidth]{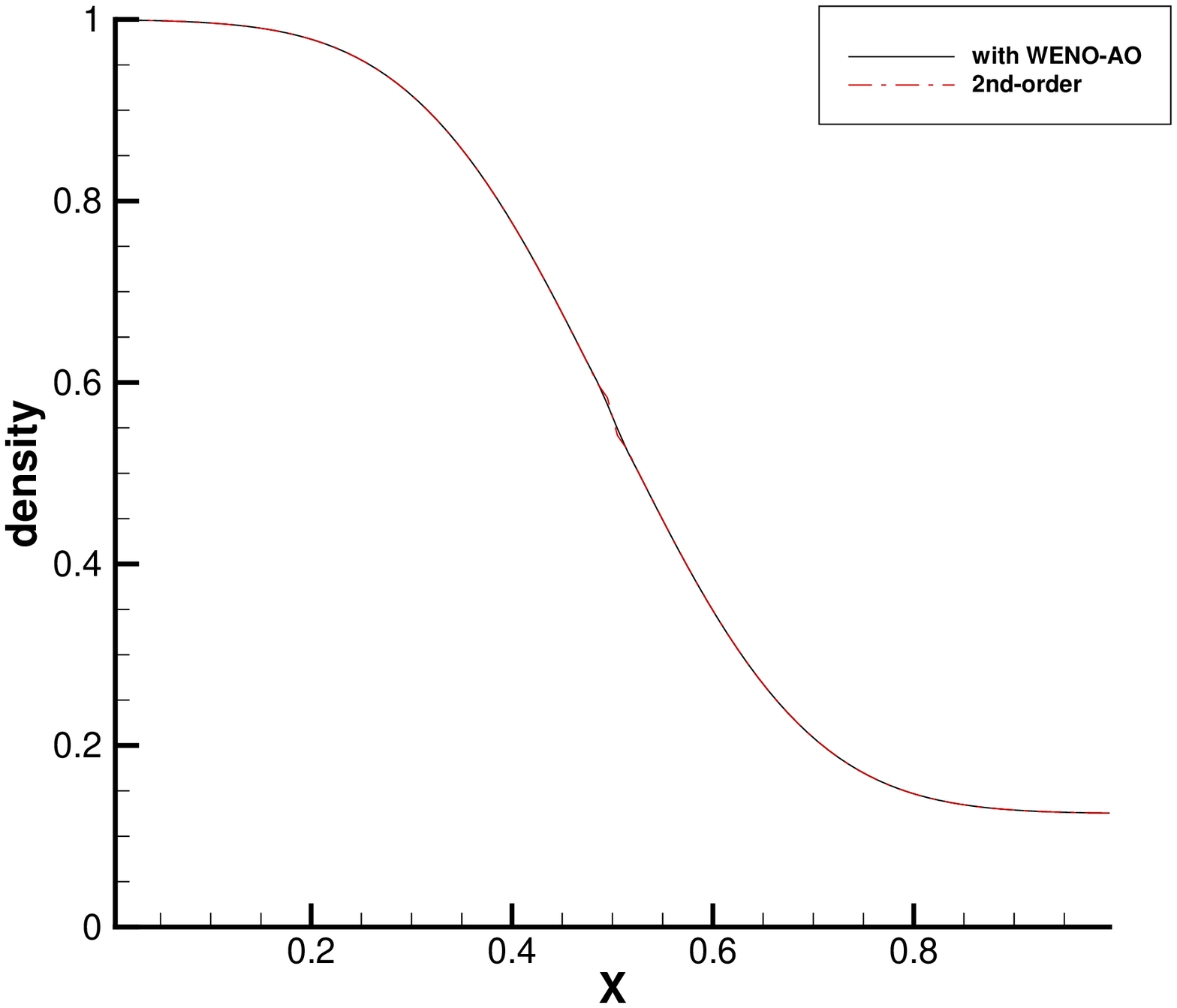}
	\includegraphics[width=0.5\textwidth]{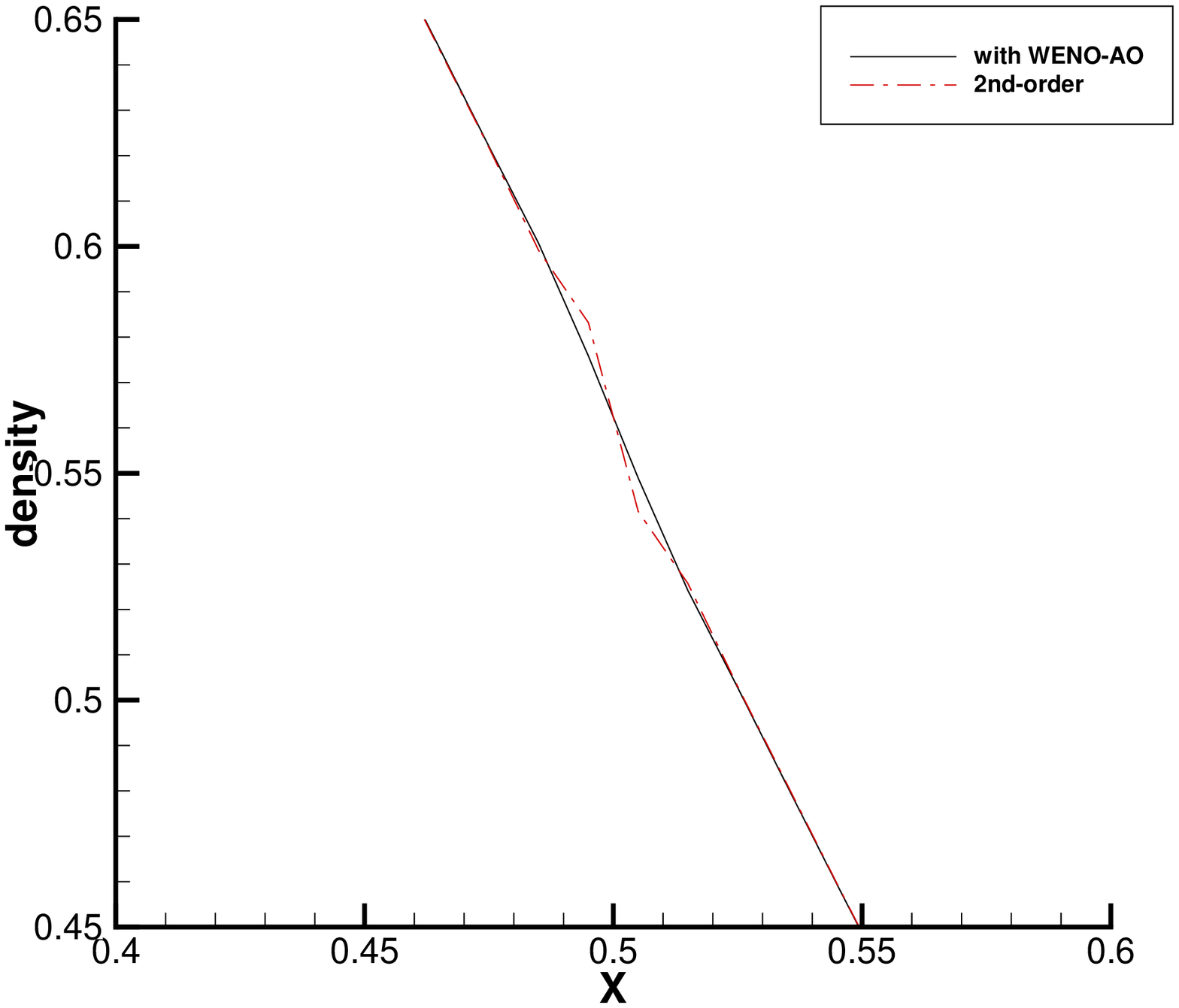}
	\caption{Sod shock tube: the density distribution and local enlargements at Kn = 10} 
	\label{Fig:sod_10}

\end{figure}
Figure~\ref{Fig:sod_1e-05} shows the density at Kn = $10^{-5}$, where the flow is in continuum regime, and it is compared with exact solution of the Euler equations calculated in \cite{Sod_euler}. The flow field gives a rarefaction wave, a contact discontinuity, and a shock. In this regime, WENO-AO implemented UGKS provides less dissipative result, and gives better result near contact discontinuity. The oscillation found near the discontinuity could be resolved by using characteristic variables as the variables for the reconstruction. 
Figure~\ref{Fig:sod_1e-03} shows the density at Kn = $10^{-3}$, where the flow is in slip regime. The discontinuity can be observed, and results from both schemes provide indistinguishable result. In this regime, the both result show deviations from exact solution of the Euler equation since the flow field is slightly rarefied. 
Figure~\ref{Fig:sod_10} shows the density at Kn = 10, and this time the results are compared with the solution of the collisionless Boltzmann equation \cite{dugks2015}. While both original second-order UGKS and WENO-AO implemented UGKS agree with collisionless Boltzmann equation solution, WENO-AO implemented UGKS provides more smooth solution at the center.

\subsubsection{Couette flow}
The Couette flow is a steady flow that is driven by the surface shearing of two infinite and parallel plates moving oppositely along their own planes. The global Knudsen number is defined as $Kn=l_{HS}/h$, where $l_{HS}$ is the mean free path based on hard sphere model, and $h$ is the distance between plates. 

\begin{figure}[htb!]
	\centering
	\includegraphics[width=0.5\textwidth]{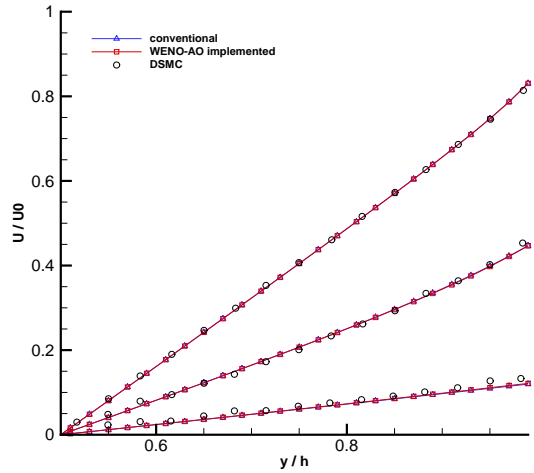}
	\caption{Couette flows: velocity profile comparison at 0.2/$\sqrt{\pi}$, 2/$\sqrt{\pi}$, and 20/$\sqrt{\pi}$ between original second-order UGKS, WENO-AO implemented UGKS and information preserving (IP) method.} 
	\label{Fig:Couette}
\end{figure}

Three Knudsen numbers are considered: 0.2/$\sqrt{\pi}$, 2/$\sqrt{\pi}$, and 20/$\sqrt{\pi}$. Physical domain is discretized with 50 cells. Figure~\ref{Fig:Couette} compares the velocity profiles given by UGKS and WENO-AO implemented UGKS with information preserving (IP) DSMC results \cite{Couette_DSMC}. $80 \times 80 $ uniform discrete velocity points are used for all cases. 
\\

The high-order UGKS could recover the non-equilibrium results as the original second-order UGKS in the transition regime.
Numerical solutions from both schemes show good agreement with the IP-DSMC data. 

\begin{figure}[htb!]
	\includegraphics[width=0.5\textwidth]{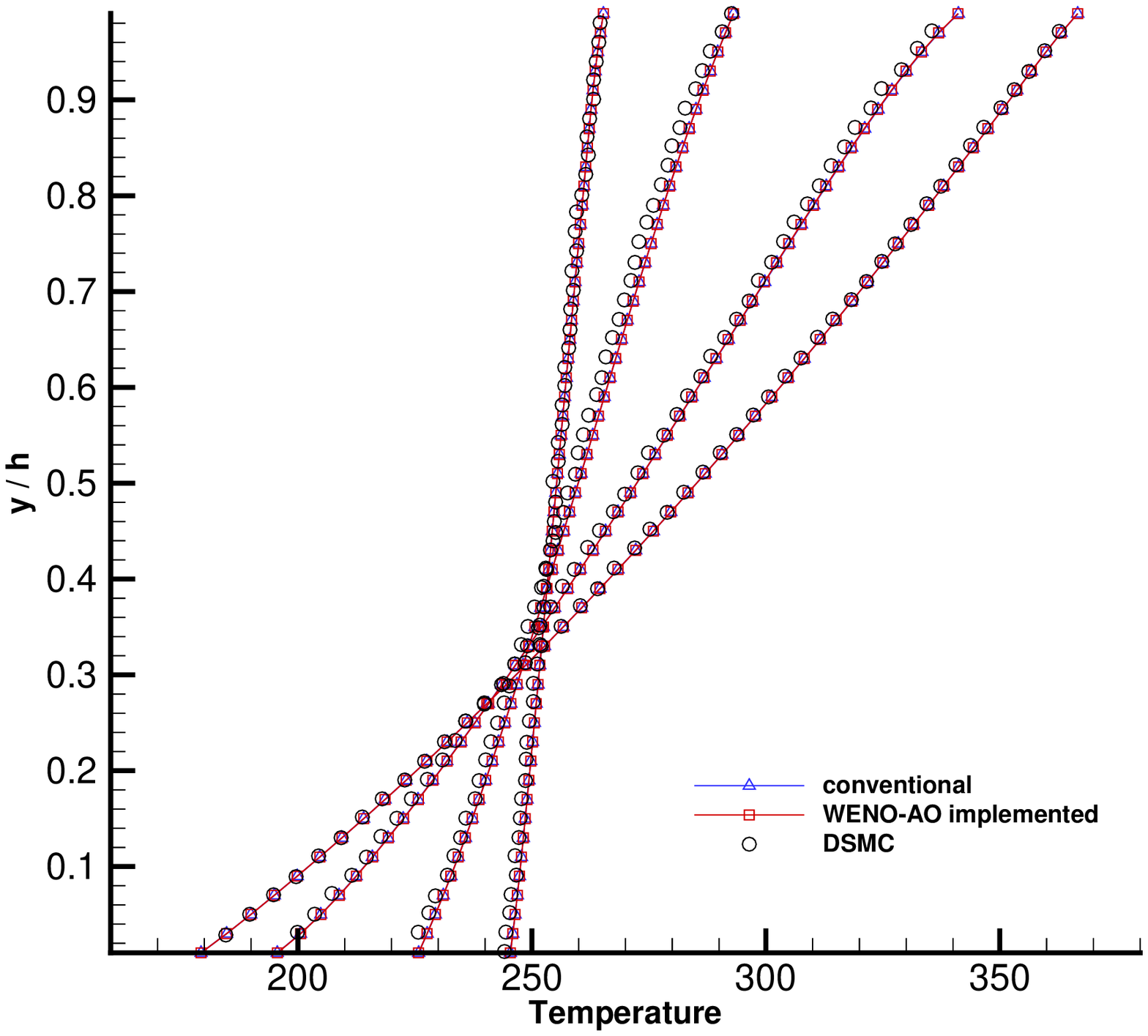}
	\includegraphics[width=0.5\textwidth]{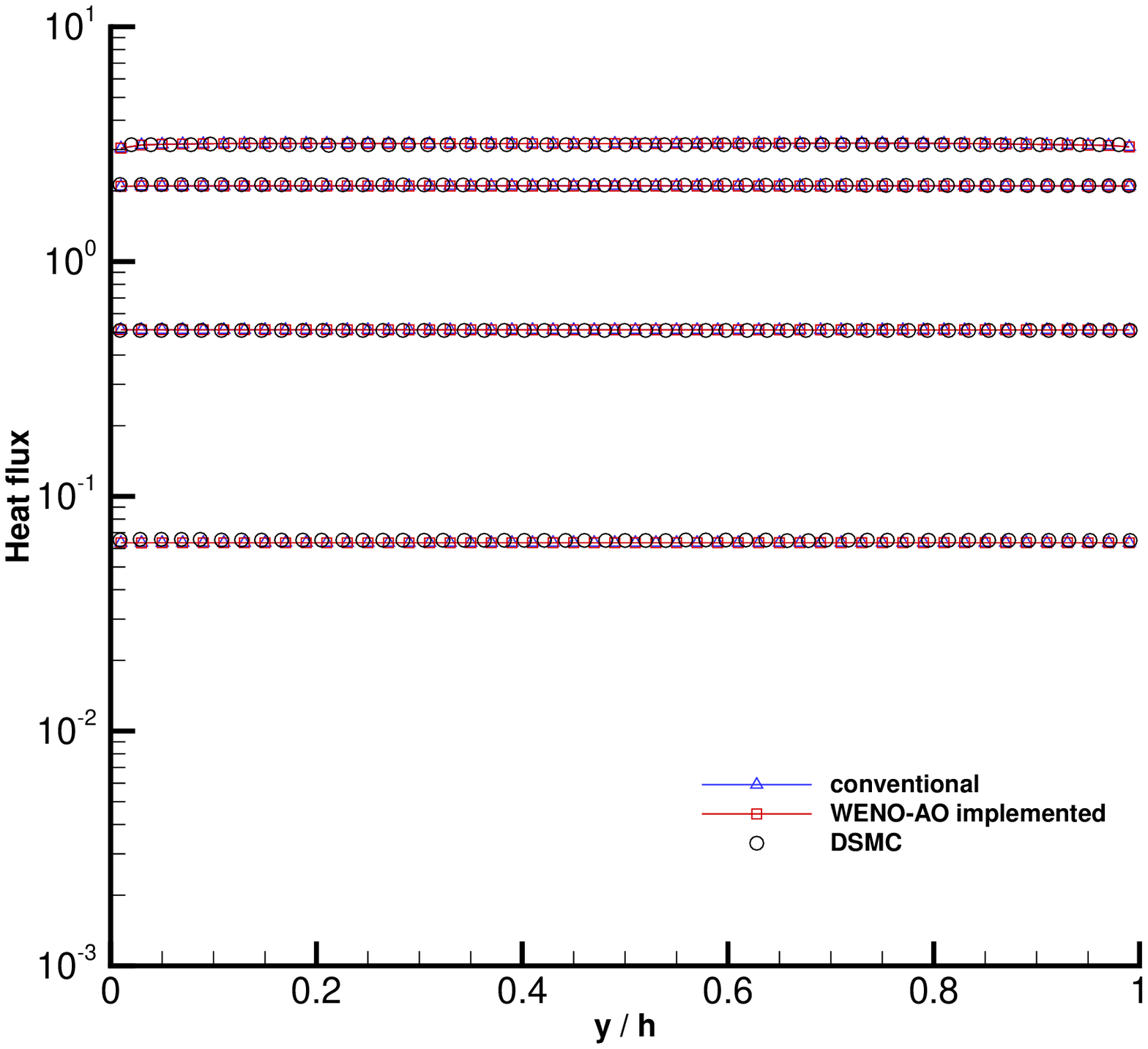}
	\caption{Thermal Couette flows: temperature(left) and heat flux(right) profile comparison at Kn = 0.01, 0.1, 1 and 10 between original second-order UGKS, WENO-AO implemented UGKS and DSMC data} 
	\label{Fig:thermalCouette}
\end{figure}

Thermal Couette flow test is a simple heat conduction problem, which is usually computed for validation of rarefied flow simulations. Two stationary parallel walls with different temperature are located. The up and down surfaces are maintained at temperature of 173K and 373K separately. The inner domain consist of monatomic argon gas at different Knudsen numbers: 0.001, 0.01, 0.1, 1, and 10. The physical domain is discretized with 50 cells, and $ 100 \times 100 $ uniform discrete velocity points are used. To validate the result, DSMC data from \cite{thermalCouette_DSMC} is plotted together in Figure~\ref{Fig:thermalCouette}. 
Good agreement with the DSMC results has also been obtained from the original second-order UGKS and WENO-AO UGKS for temperature and heat flux profiles in this heat conduction problem.

\subsubsection{Oscillatory Couette flow}
The oscillatory Couette flow is unsteady rarefied gas flow between two infinite parallel plates. The bottom plate has periodic oscillation in lateral direction, and top plate is stationary. Both plates are isothermal wall with 273K. Zhang \cite{oscillatoryCouette} introduced two parameters which characterizes the flow field. One is the rarefaction parameter $\delta$, and the other is oscillation parameter $\theta$. Each parameter is defined as following
\begin{equation}\label{osc_Couette_parameters}
	\delta = \frac{p_0h}{\mu \nu_0}, ~~ \theta = \frac{p_0}{\mu \omega_0}, 
\end{equation}
where $p_0 = n_0 k_B T_0$ is the equilibrium pressure  of the gas, with $\mu_0$ is dynamic viscosity at $T_0$. The rarefaction parameter is related to global Knudsen number. For hard sphere model, two parameters are related by $\delta = 0.5\sqrt{\pi}/\mathrm{Kn}$. The oscillation parameter is related to the frequency ratio, which is defined as the ratio of intermolecular collision frequency $p_0/\mu$ to the oscillation frequency of the plate $\omega_0$. While Stokes number is often used to describe oscillation parameter, different oscillation parameter is used to describe the non-equilibrium effect on the time scale caused by oscillation. Since Stokes number is used to describe the balance between the unsteady and the viscous effects, it is not sufficient to include non-equilibrium effect in time scale. The oscillation parameter is related to Stokes number as
\begin{equation}\label{osc_Couette_Stk}
Stk = \sqrt{\frac{\omega_0h^2}{\nu}}=\sqrt{\frac{2}{\theta}}\delta, ~~ \theta = 2 (\frac{\delta}{Stk})^2
\end{equation}
where $\nu$ is the kinematic viscosity of the gas. By using two parameters, both of the spatial and temporal rarefaction can be evaluated. When $\delta$ is large enough, the characteristic length is much larger than mean free path of the molecules. In contrast, when $\delta$ is close to zero, the characteristic length is small, and it is comparable to the mean free path. When $\theta$ is large, the oscillation frequency is low, which results quasi-stationary flow. In contrast, small $\theta$ will give high oscillation frequency, which results almost no intermolecular collision during one oscillation period. Table~\ref{Table:oscCouette} illustrates the corresponding Kn and $Stk$ for each case. 

The oscillating Couette test with different rarefaction and oscillation parameters are computed. All tests are discretized with 100 cells in physical domain, and different velocity space discretization is applied according to the rarefaction parameter. (i.e., $8\times8$ Gaussian-Hermite velocity space is used for $\delta$ = 100 and 1000, and $28\times28$ Gaussian-Hermite velocity space is used for $\delta$ = 10). The velocity profile result are obtained at $t = 0.25 t_0, 0.5 t_0, 0.75 t_0$ and $t_0$ for each case, where $t_0 = 2\pi/\omega$, which is the period of oscillating plate. The solutions are in steady periodic state with the average relative difference between the two results in two successive periods is less than the residual. 
\\

\begin{table}[]
	\centering
	\caption{oscillating Couette flow: Kn and $Stk$ for corresponding $\delta$ and $\theta$.}
	\label{Table:oscCouette}
	\begin{tabular}{||c|c||c|c||}
		\hline
		rarefaction parameter, $\delta$ & oscillation parameter, $\theta$ & Knudsen number       & Stokes number  \\ \hline
		1000     & 10000    & 0.000886 & 14.142 \\ \hline
		1000     & 1000     & 0.000886 & 44.721 \\ \hline
		100      & 100      & 0.00886  & 14.142 \\ \hline
		10       & 1        & 0.0886   & 14.142 \\ \hline
	\end{tabular}
\end{table}

\begin{figure}[htb!]
	\includegraphics[width=0.5\textwidth]{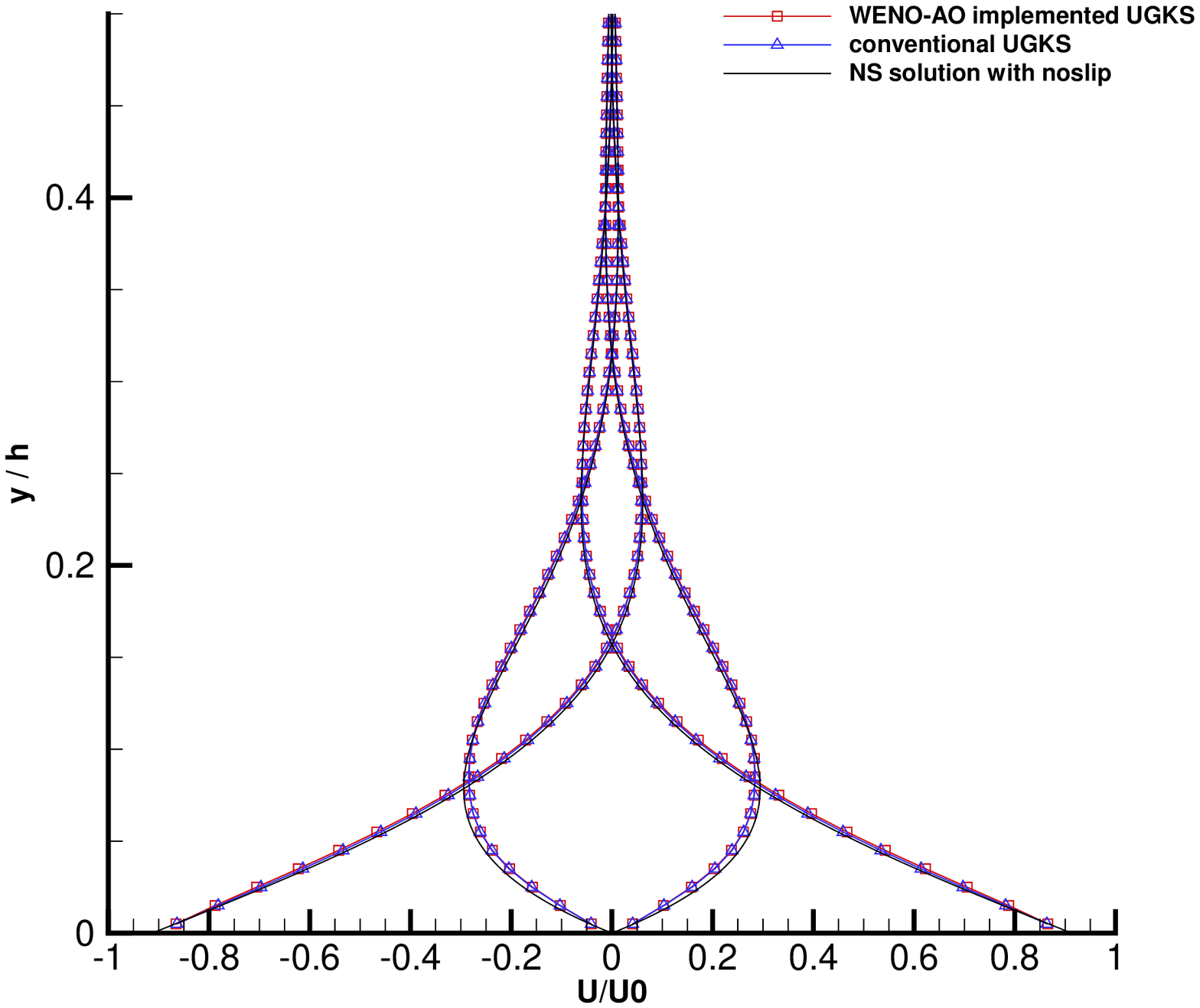}
	\includegraphics[width=0.5\textwidth]{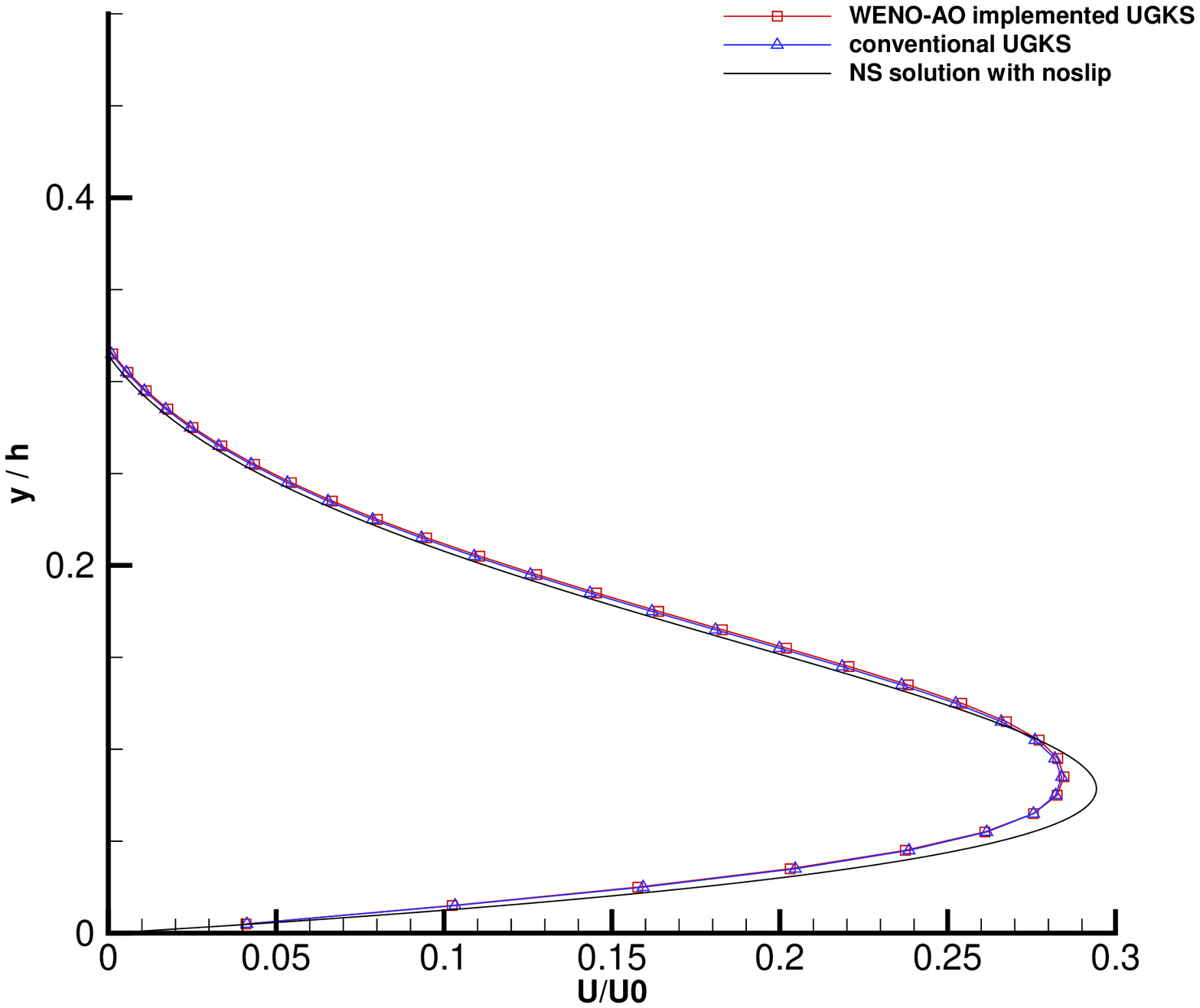}
	\caption{oscillating Couette flows: horizontal velocity profile comparison at $\delta$ = $10^{3}$ and $\theta$ = $10^{4}$ between no-slip Navier-Stokes solution, original second-order UGKS and WENO-AO implemented UGKS, and its local enlargement at $t = 0.25 t_0$.} 
	\label{Fig:oscCouette1000x10000}
\end{figure}

\begin{figure}[htb!]
	\includegraphics[width=0.5\textwidth]{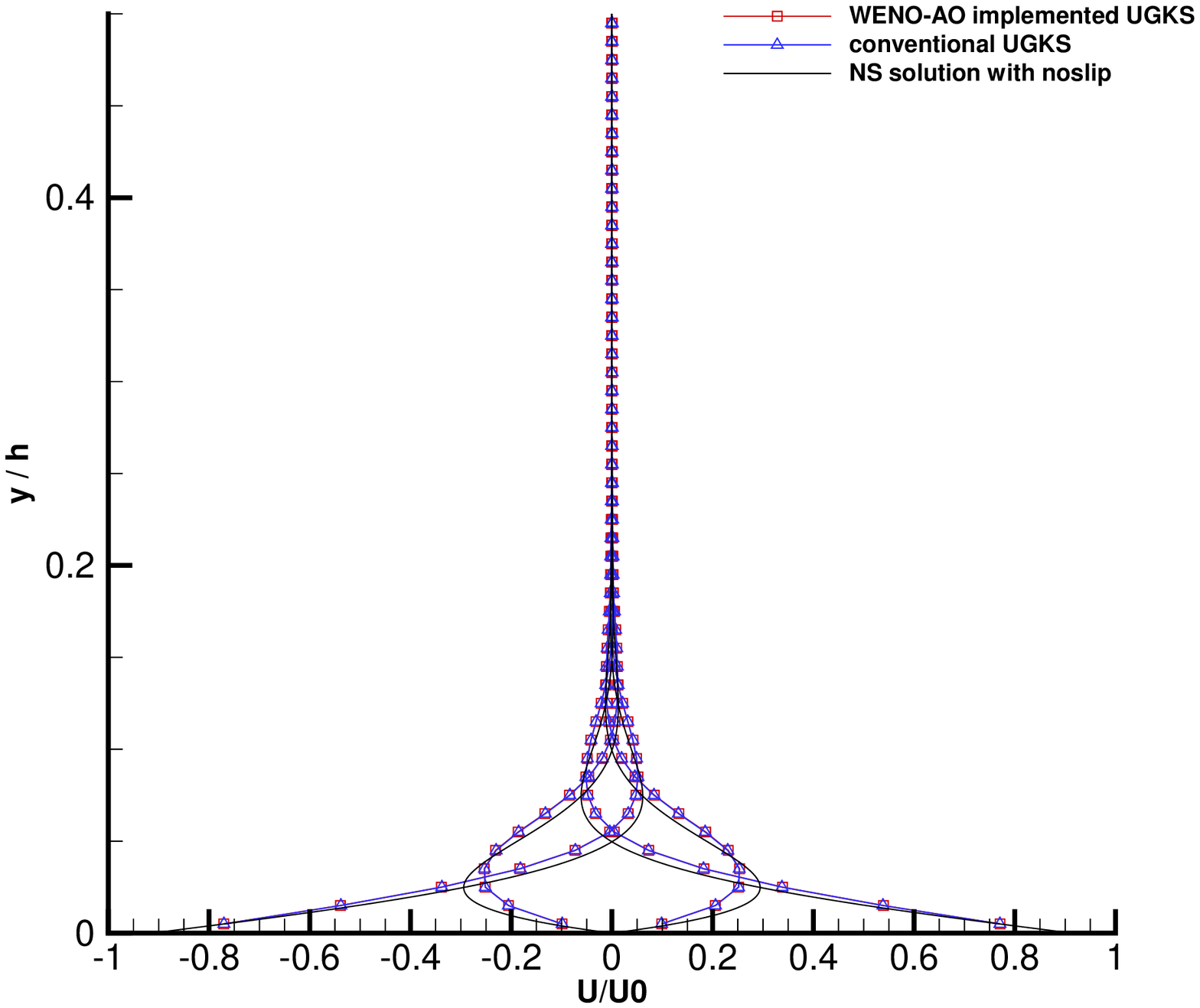}
	\includegraphics[width=0.5\textwidth]{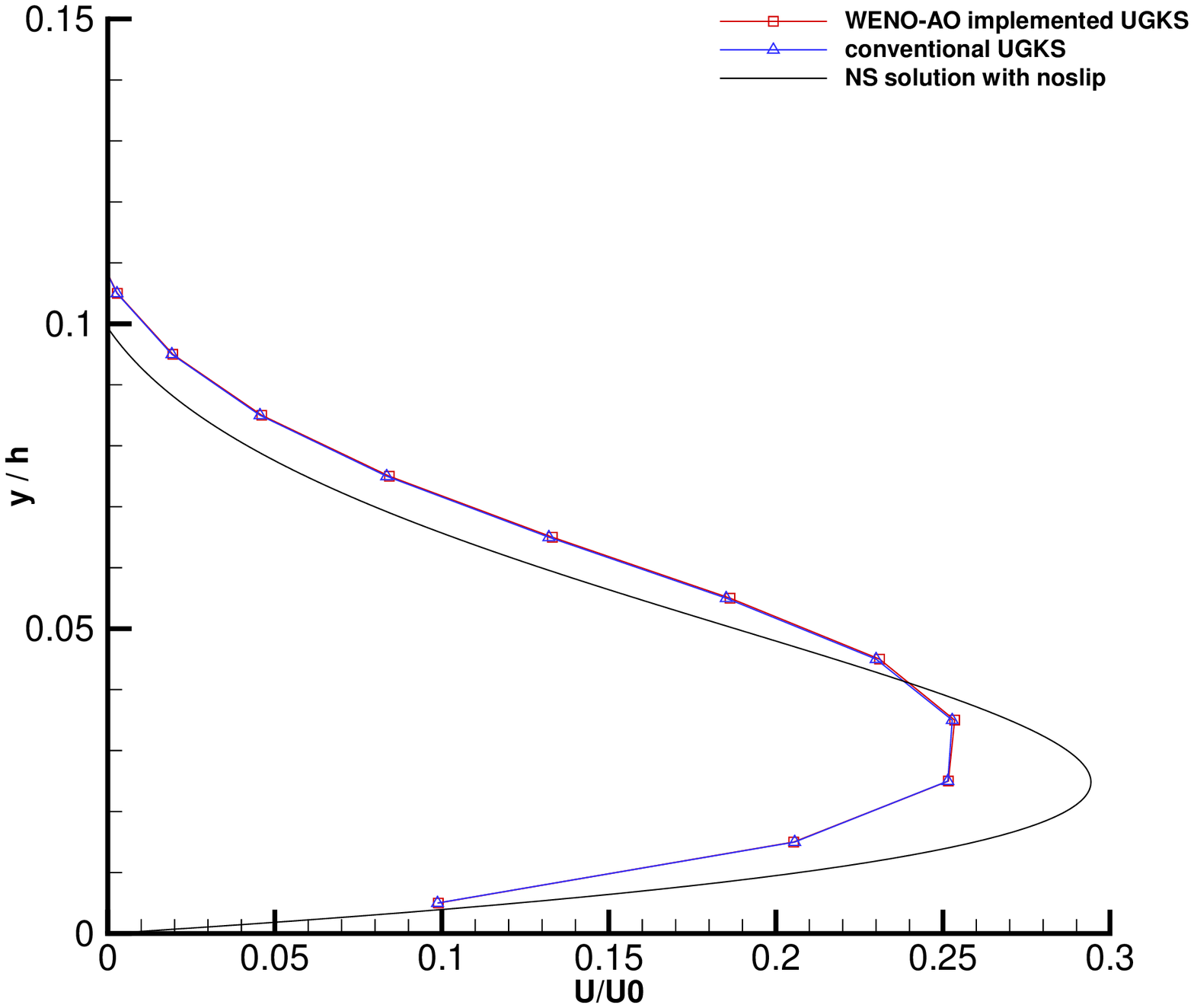}
	\caption{oscillating Couette flows: horizontal velocity profile comparison at $\delta$ = $10^{3}$ and $\theta$ = $10^{3}$ between no-slip Navier-Stokes solution, original second-order UGKS and WENO-AO implemented UGKS, and its local enlargement at $t = 0.25 t_0$.} 
	\label{Fig:oscCouette1000x1000}
\end{figure}

\begin{figure}[htb!]
	\includegraphics[width=0.5\textwidth]{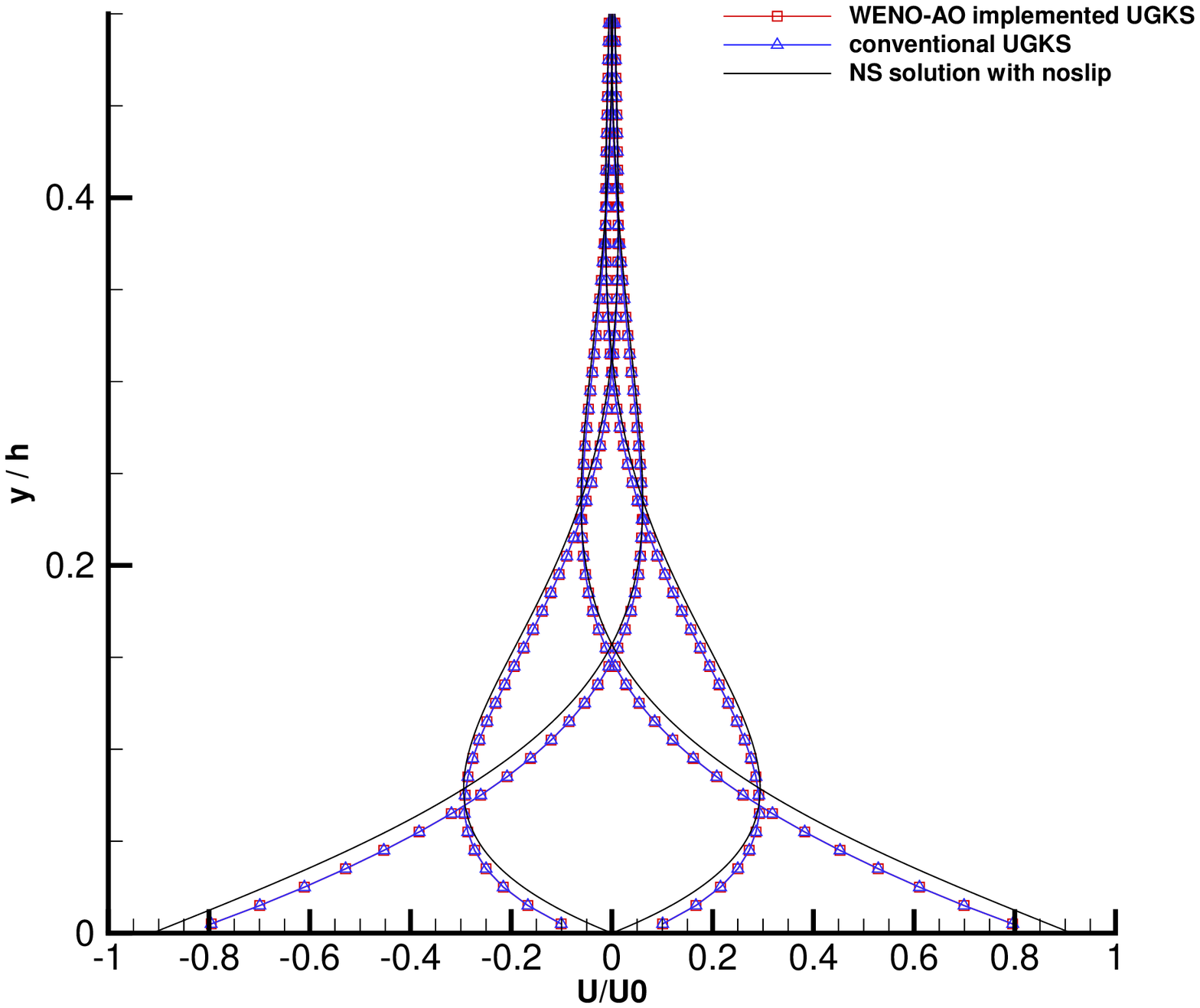}
	\includegraphics[width=0.5\textwidth]{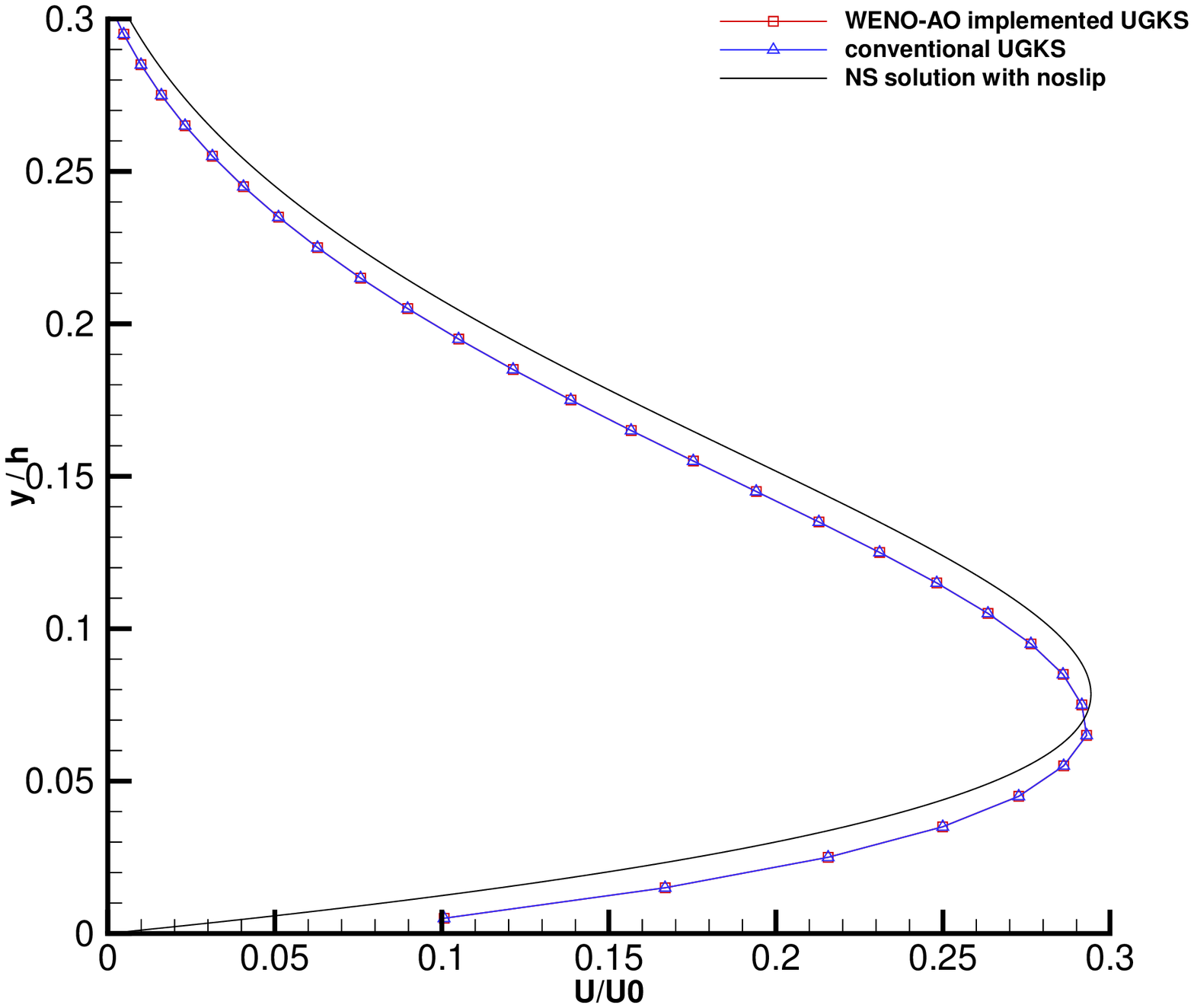}
	\caption{oscillating Couette flows: horizontal velocity profile comparison at $\delta$ = $100$ and $\theta$ = $100$ between slip Navier-Stokes solution, original second-order UGKS and WENO-AO implemented UGKS, and its local enlargement at $t = 0.25 t_0$.} 
	\label{Fig:oscCouette100x100}
\end{figure}

\begin{figure}[htb!]
	\includegraphics[width=0.5\textwidth]{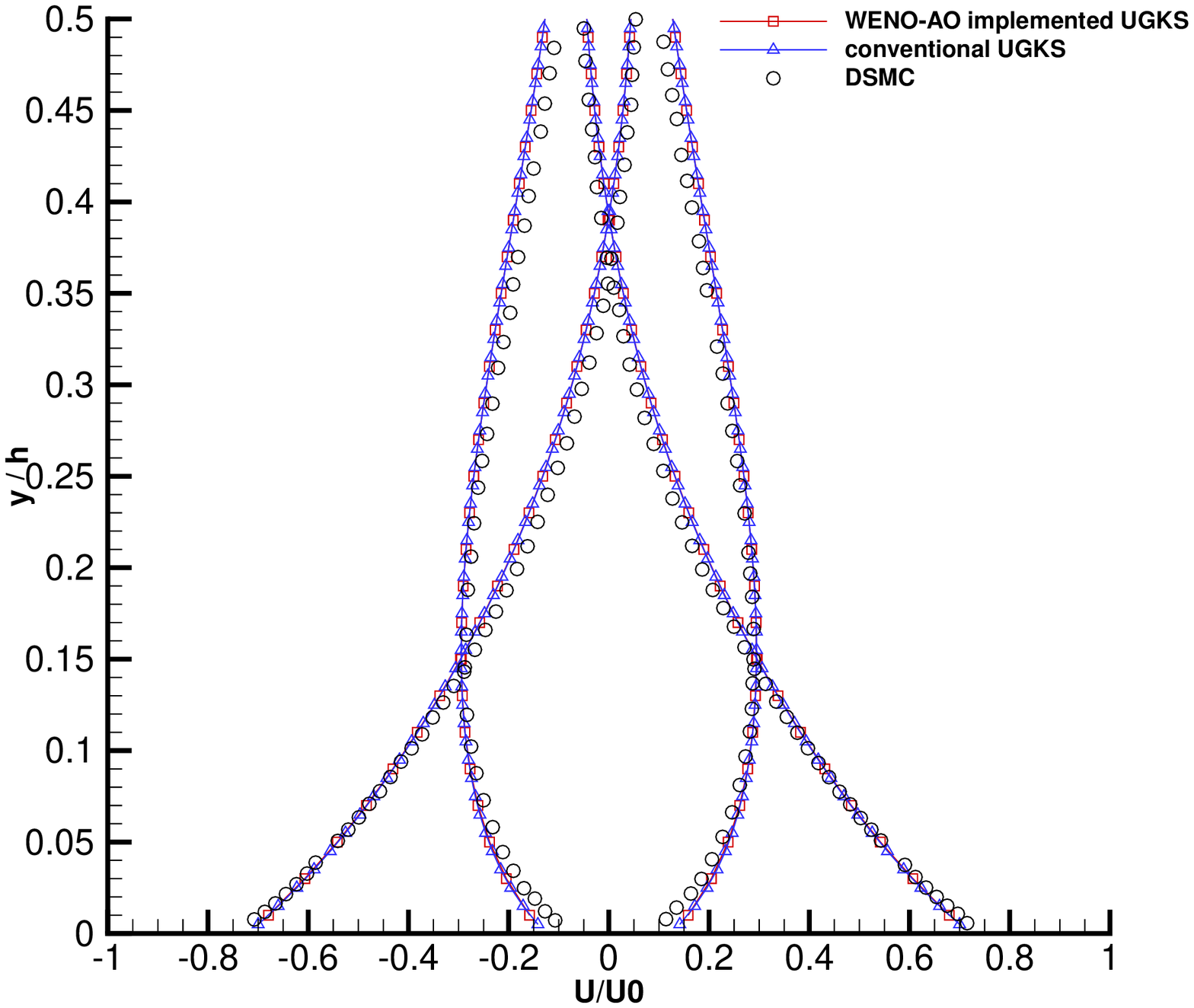}
	\includegraphics[width=0.5\textwidth]{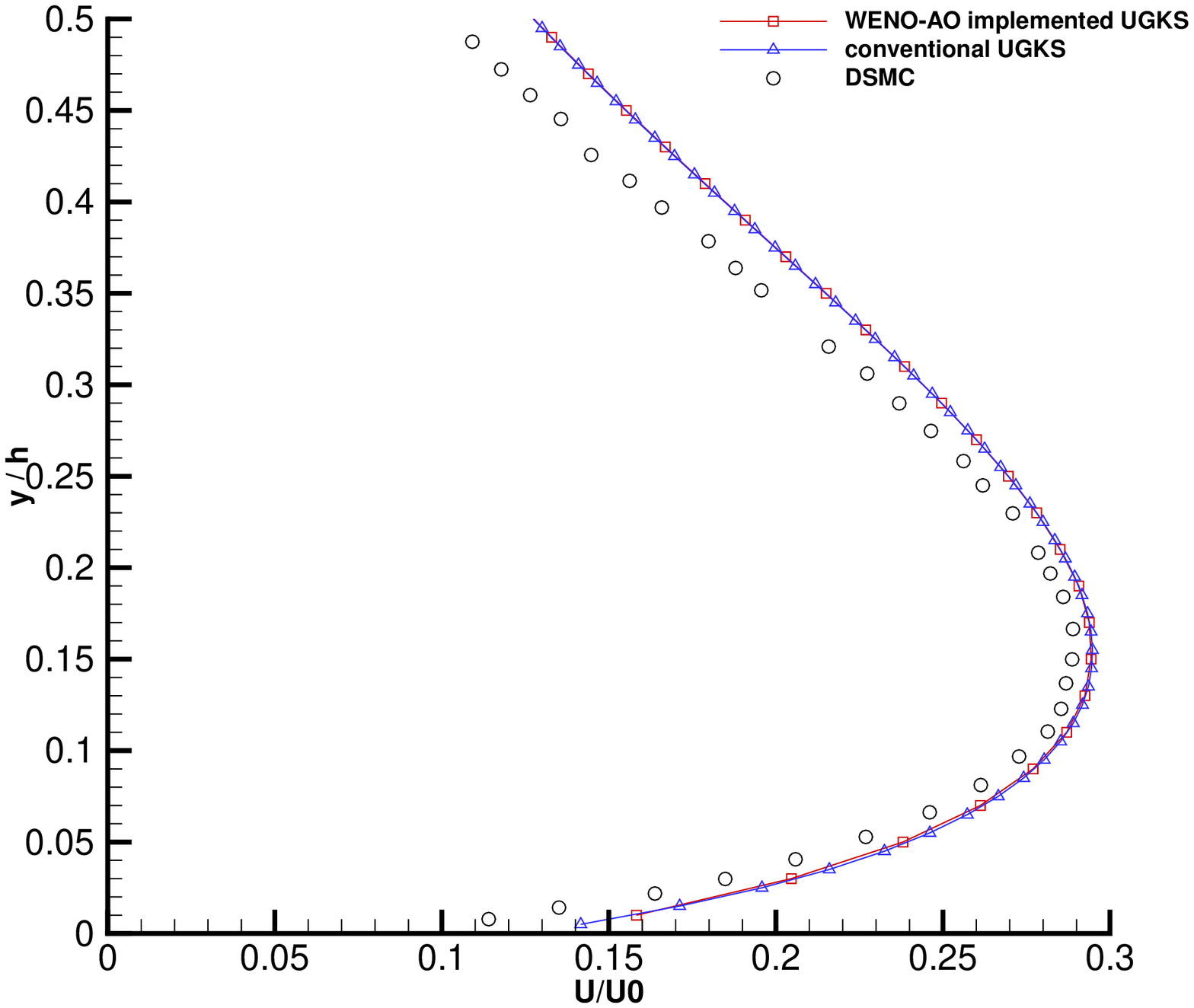}
	\caption{oscillating Couette flows: horizontal velocity profile comparison at $\delta$ = $10$ and $\theta$ = $1$ between DSMC data, original second-order UGKS and WENO-AO implemented UGKS, and its local enlargement at $t = 0.25 t_0$.} 
	\label{Fig:oscCouette10x10}
\end{figure}

Figure~\ref{Fig:oscCouette1000x10000} illustrates original second-order UGKS, WENO-AO implemented UGKS and analytical solution of incompressible Navier-Stokes equation with no-slip condition \cite{Landau_fluidmechanics} at $\delta$ = 1000 and $\theta$ = 10000. It is found that both original second-order UGKS and WENO-AO implemented UGKS provide good agreement to the Navier-Stokes solution. In Figure~\ref{Fig:oscCouette1000x1000}, $\theta$ is reduced to 1000 to investigate the results at high oscillation frequency. It can be found that WENO-AO implemented UGKS provides slightly higher peak velocity, which is closer to the reference data. Thus, WENO-AO implemented UGKS can give better description for oscillating Couette flow at higher Stokes number. 

Figure~\ref{Fig:oscCouette100x100} gives original second-order UGKS, WENO-AO implemented UGKS and Navier-Stokes solution with slip boundary condition at $\delta$ = 100 and $\theta$ = 100, which gives identical $Stk$ with Figure~\ref{Fig:oscCouette1000x10000} but different rarefaction parameter. Due to slightly rarefied flow field, Navier-Stokes solution with no-slip is not perfectly valid in slip flow regime. Thus, Navier-Stokes with slip condition \cite{oscillatoryCouette_slipbc} is used to validate the results at $\delta$ = 100. The result from original second-order and WENO-AO implemented UGKS are almost identical to each other and both provide great agreement with the solution of Navier-Stokes equation with slip condition. It is observed that WENO-AO implemented UGKS can recover original second-order UGKS in slip flow regime. 

To validate UGKS solution in transition regime, Figure~\ref{Fig:oscCouette10x10} gives UGKS solution at $\delta$ = 10 and $\theta$ = 10 with DSMC data from \cite{oscillatoryCouette_DSMC}. The simulation is not conducted in perfectly same condition, but they are close to each other. (i.e., DSMC data is conducted with Kn = 0.1 and $Stk$ = 5.0 while UGKS solution is conducted with Kn = 0.0886 and $Stk$ = 4.4721.)  As expected, the result from original second-order and WENO-AO implemented UGKS provides are very close to DSMC solution. The small difference could be caused by difference in Kn and $Stk$. Thus, UGKS can compute flow in transition regime, and WENO-AO implemented UGKS can recover original second-order UGKS well.

\subsection{2-D test cases}
\subsubsection{2-D sine wave accuracy test}
The advection of density perbutation is also tested in two-dimensions. The physical domain is set as [0,2]$\times$[0,2] with $N\times N$ uniform mesh cells, and initial condition is given as following
\begin{equation}\label{2d_accuracy_initial}
	\rho(x, y) = 1+0.2\sin(\pi (x+y)),\ U(x,y) = 1, \ p(x,y) = 1,  
\end{equation}
When the periodic boundary condition applied at each end, the exact solution is 
\begin{equation}\label{2d_accuracy_analytic}
	\rho(x,y,t) = 1+0.2\sin(\pi (x+y-t)), \ U(x,y,t) = 1,\ p(x,y,t)=1, 
\end{equation}
The test details are same as 1-D sine wave accuracy test. 
The $L^1$, $L^2$, and $L^\infty$ errors and the corresponding orders of 2-D sine wave accuracy test at $t = 2$ are tabulated in Table~\ref{Table:2D-accuracy-second} and Table~\ref{Table:2D-accuracy-weno}. Like 1-D sine wave accuracy test, the original second-order UGKS gives lower-order of accuracy in Table~\ref{Table:2D-accuracy-second}, WENO5-AO implemented UGKS provides fifth-order of accuracy in 2-D sine wave accuracy test as expected in Table~\ref{Table:2D-accuracy-weno}. 

\begin{table}[]
	\centering
	\caption{Accuracy test for the 2-D sine wave propagation by the original second-order UGKS with smooth solver.}
	\label{Table:2D-accuracy-second}
	\begin{tabular}{||c|cc|cc|cc||} \hline
		mesh length	&$L^1$ error  &Order  &$L^2$ error  &Order &$L^\infty$ error  &Order  \\ \hline
		1/5			&2.09837E-01  &  &1.67617E-01  &  &1.61167E-01  &  \\
		1/10		&1.08654E-01  &0.949528   &8.87945E-02  &0.916626   &9.12853E-02  &0.820102   \\
		1/20		&3.25831E-02  &1.737546    &2.76347E-02  &1.683989    &3.54480E-02  &1.364678    \\ 
		1/40		&1.26228E-02  &1.368092    &1.02064E-02  &1.437007    &1.35667E-02  &1.385634    \\
		1/80		&3.39794E-03  &1.893300    &3.08077E-03  &1.728111    &5.08542E-03  &1.415631    \\ \hline
	\end{tabular}
\end{table}

\begin{table}[]
	\centering
	\caption{Accuracy test for the 2-D sine wave propagation by the WENO5-AO implemented UGKS with smooth solver. The linear weights of $\gamma_{Hi} = 0.85$, $\gamma_{Lo} = 0.85$ used.}
	\label{Table:2D-accuracy-weno}
	\begin{tabular}{||c|cc|cc|cc||} \hline
		mesh length	&$L^1$ error  &Order  &$L^2$ error  &Order &$L^\infty$ error  &Order  \\ \hline
		1/5			&9.99273E-02  &  &7.78269E-02  &  &7.69734E-02  &  \\
		1/10		&3.70838E-03  &4.752018    &2.85923E-03  &4.766570    &2.87722E-03  &4.741613   \\
		1/20		&1.17503E-04  &4.980020     &9.21434E-05  &4.955602     &9.47986E-05  &4.923666      \\ 
		1/40		&3.72876E-06  &4.977858     &2.92670E-06  &4.976534     &3.04737E-06  &4.959229     \\
		1/80		&1.19422E-07  &4.964555     &9.37650E-08  &4.964082    &9.73221E-08  &4.968653    \\ \hline
	\end{tabular}
\end{table}

\subsubsection{Cavity flow}
\begin{figure}[H]
	\centering
	\includegraphics[width=0.5\textwidth]{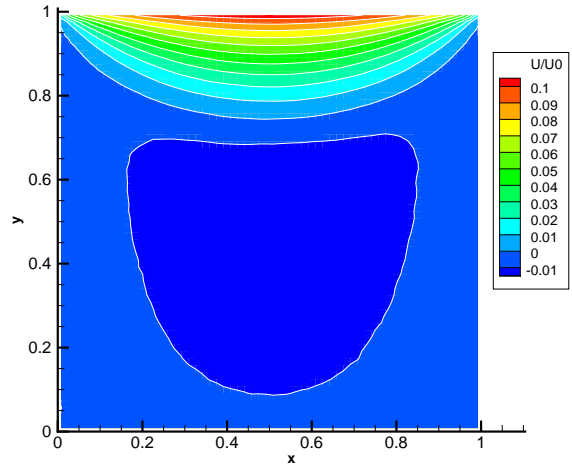}
	\caption{cavity flow horizontal velocity distributions at Kn = 0.075. Black lines: reference data from UGKS with fine mesh, white lines: DSMC data.} 
	\label{Fig:cavityDSMC}
\end{figure}

\begin{figure}[H]
	\includegraphics[width=0.5\textwidth]{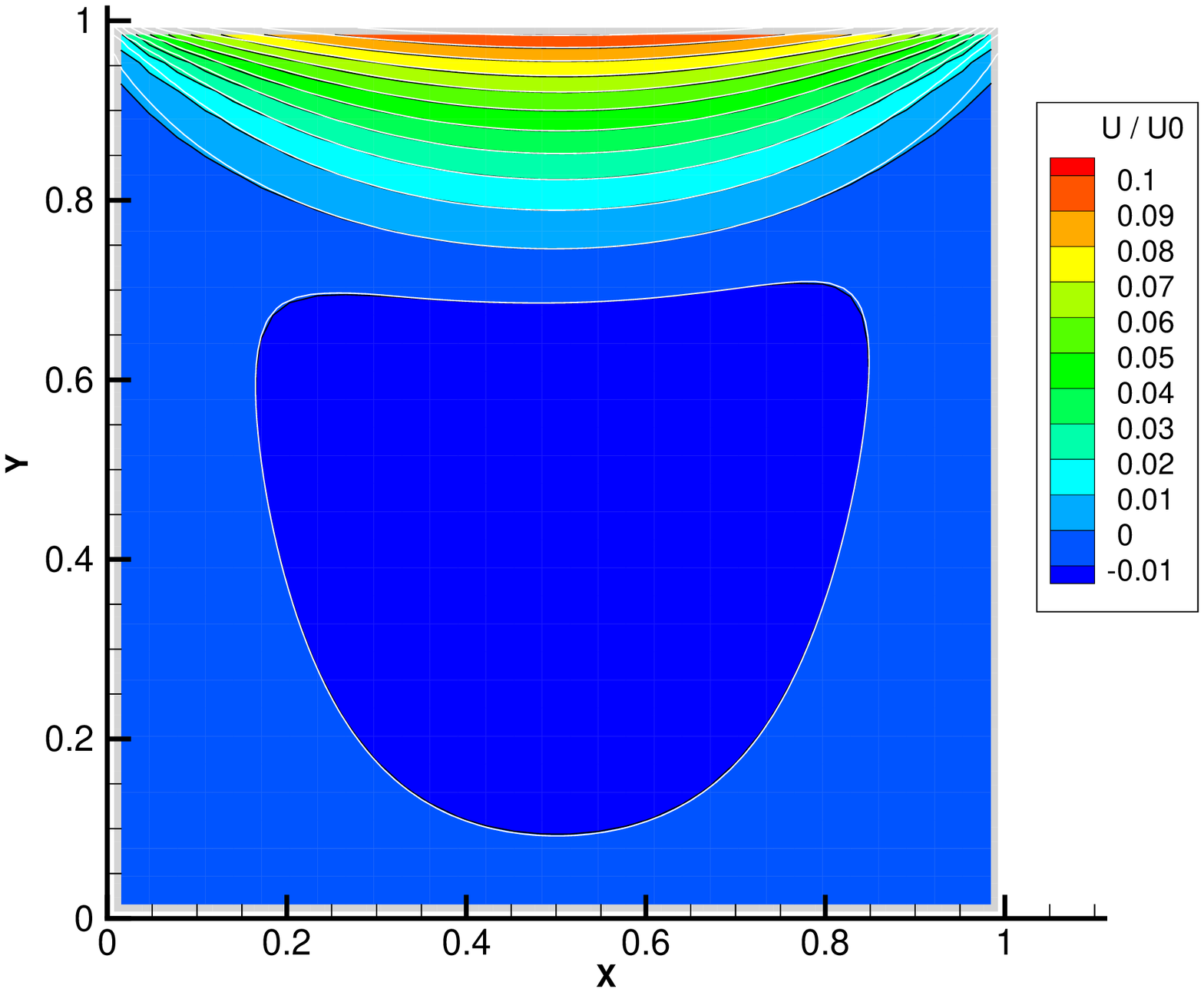}
	\includegraphics[width=0.5\textwidth]{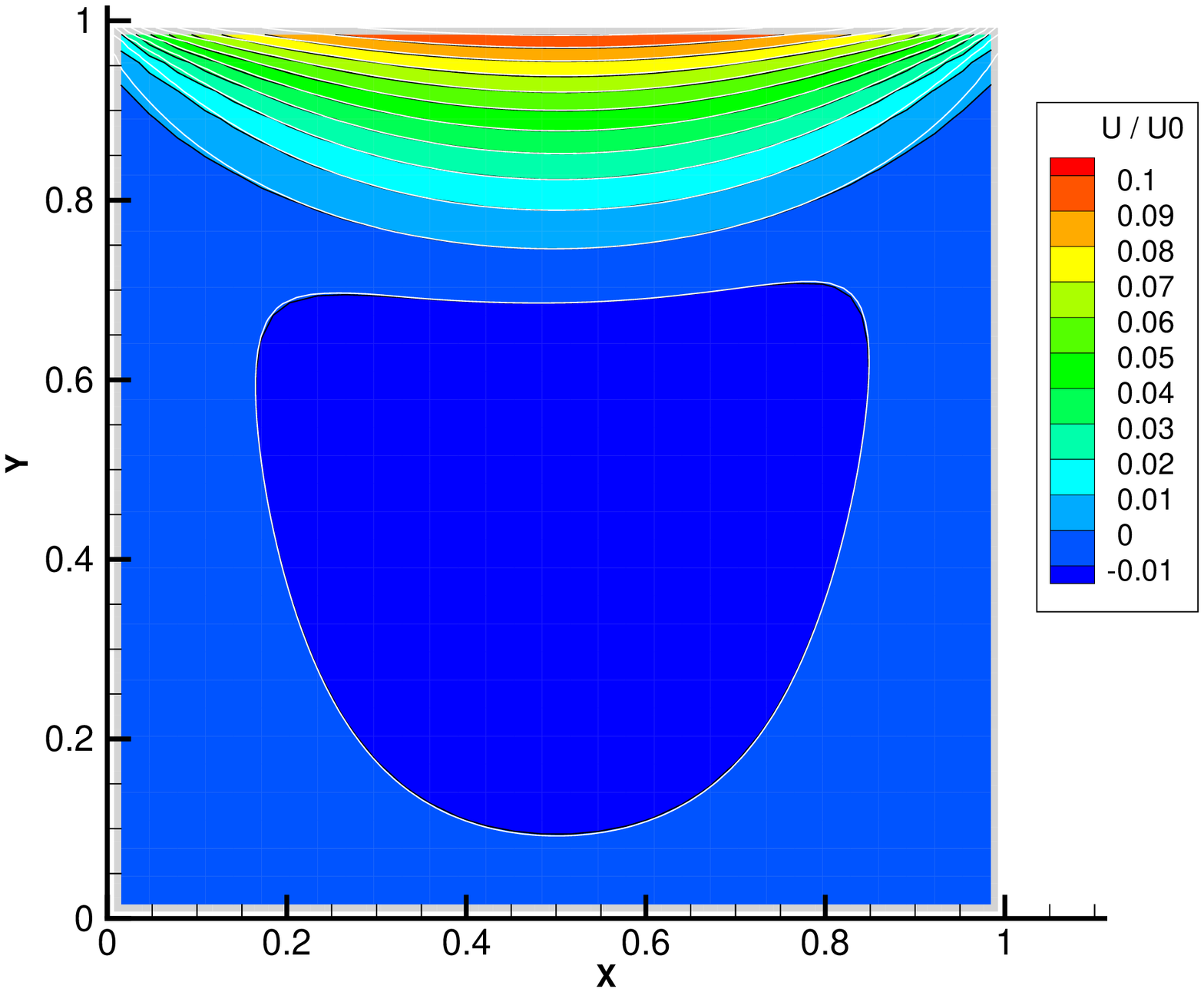}
	\caption{cavity flow horizontal velocity distributions at Kn = 0.075. Black lines: original second-order UGKS(left), WENO-AO implemented UGKS(right), white lines: reference result from original second-order UGKS on a finer mesh.} 
	\label{Fig:cavity075}
\end{figure}

\begin{figure}[htb!]
	\includegraphics[width=0.5\textwidth]{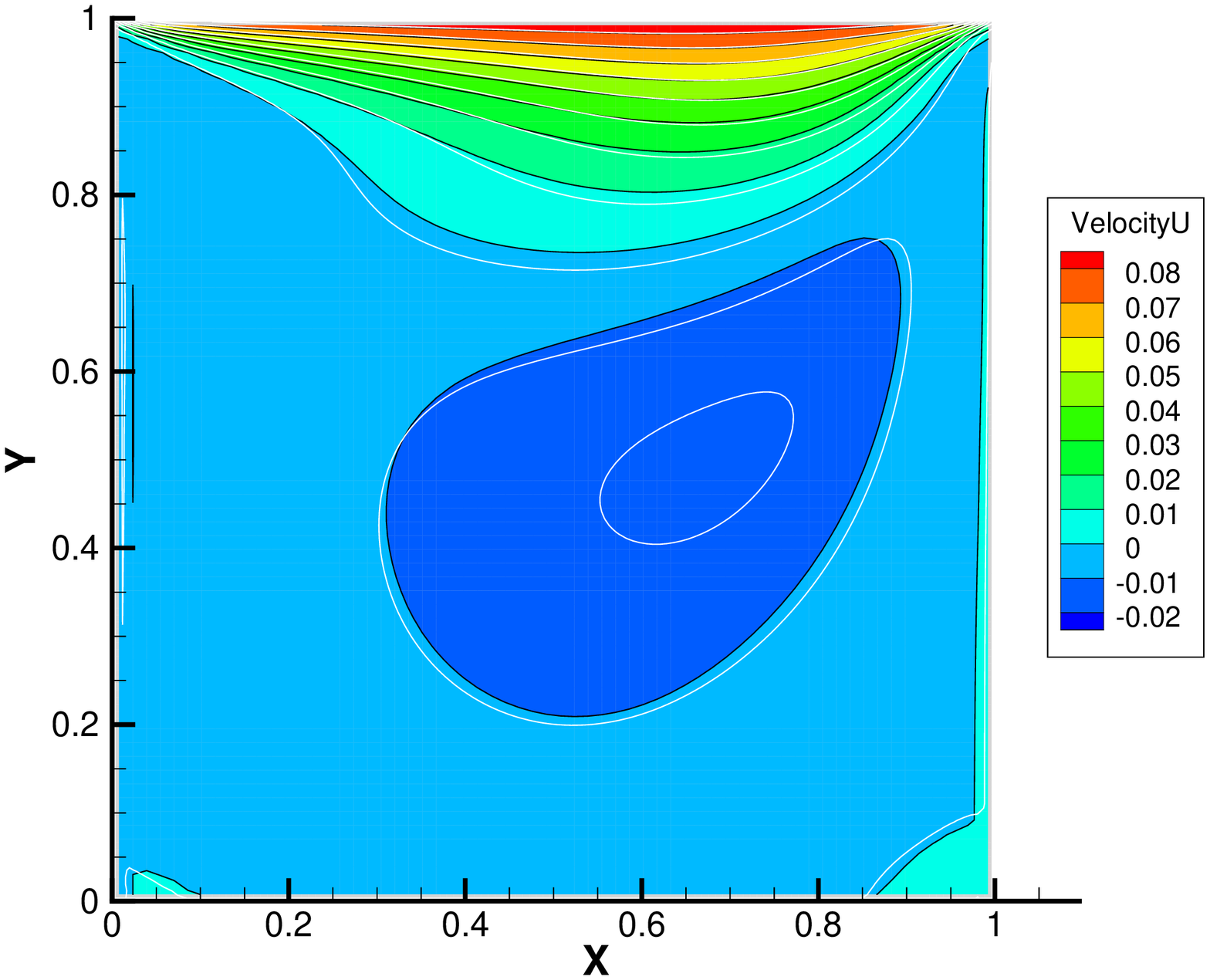}
	\includegraphics[width=0.5\textwidth]{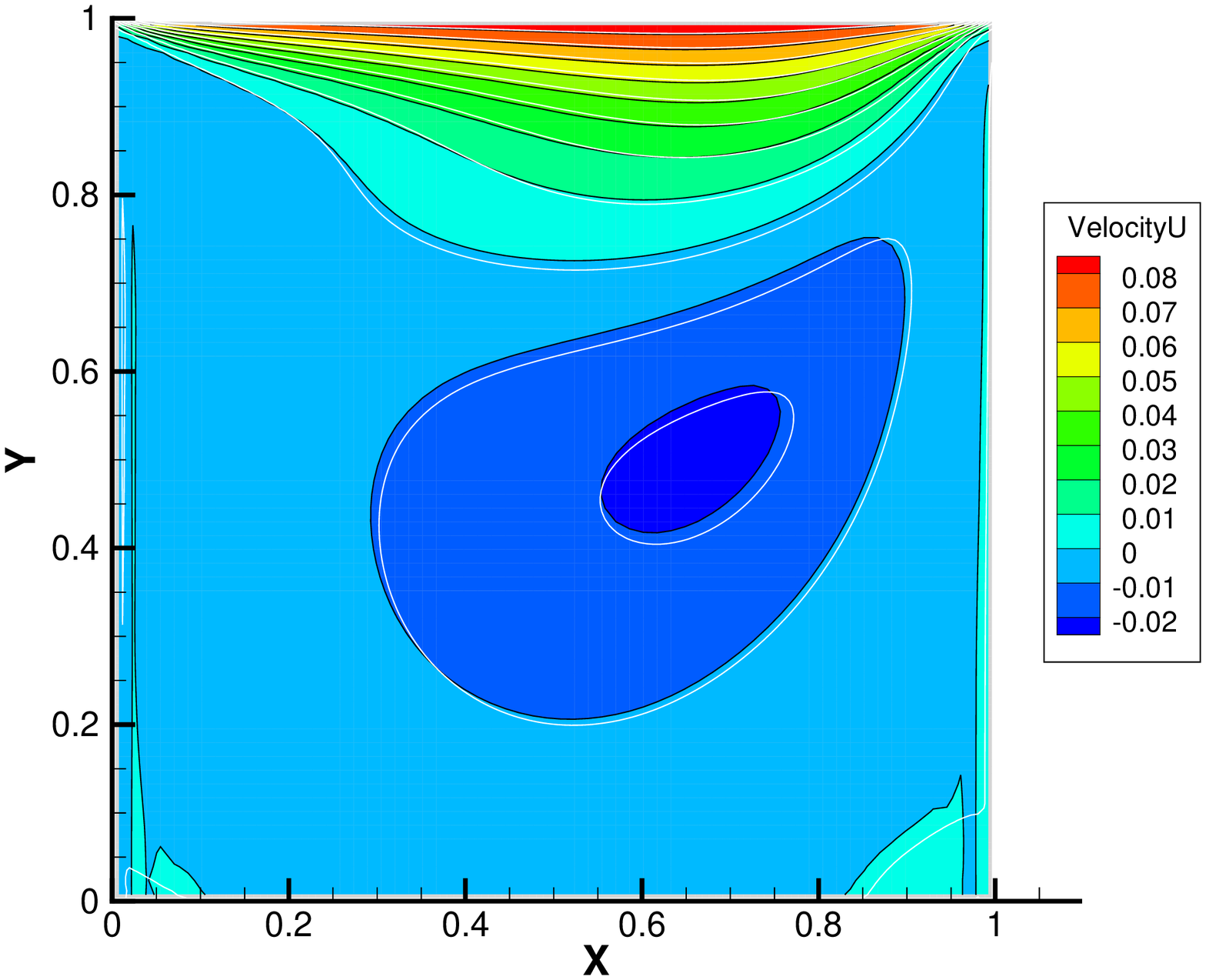}
	\caption{cavity flow horizontal velocity distributions at Kn = $10^{-3}$. Black lines: original second-order UGKS(left), WENO-AO implemented UGKS(right), white lines: reference result from original second-order UGKS with double mesh number(both).} 
	\label{Fig:cavity1e-03}
\end{figure}

The lid-driven cavity flow is a well known two-dimensional test case which consists of three isothermal stationary walls and one moving isothermal wall on the top with constant velocity. The monatmoic gas are filled inside the cavity, and the cases at two different Knudsen numbers, (i.e., Kn = 0.075 and 0.001), are considered for the calculations using the original second-order and WENO-AO implemented UGKS. To observe the difference between two schemes clearly, the coarse mesh is used to compare. The physical domain is discretized by $64\times64$ for Kn = 0.001, and $32\times32$ for Kn = 0.075. The $8\times8$ Gaussian-Hermite velocity space is used for Kn = $10^{-3}$, and $28\times28$ Gaussian-Hermite velocity space is used for Kn = 0.075. The reference is obtained from original second-order UGKS on a finer mesh with double of cells in physical domain. 

To validate the UGKS result, the reference UGKS data is compared with DSMC data \cite{cavity_DSMC}. Figure~\ref{Fig:cavityDSMC} shows that the reference data (i.e., result from original second-order UGKS with the fine mesh) agrees with the DSMC data at Kn = 0.075 and Figure~\ref{Fig:cavity075} shows that both results are almost identical to each other and agree with reference data. At lower Knudsen number, according to Figure~\ref{Fig:cavity1e-03}, WENO-AO implemented UGKS gives better agreement to the reference data. While original second-order UGKS could not describe peak negative velocity near the center of the cavity, WENO-AO implemented UGKS could provide close horizontal velocity contour to the reference data.

\subsubsection{Oscillatory cavity flow}
\begin{figure}[H]
	\includegraphics[width=0.5\textwidth]{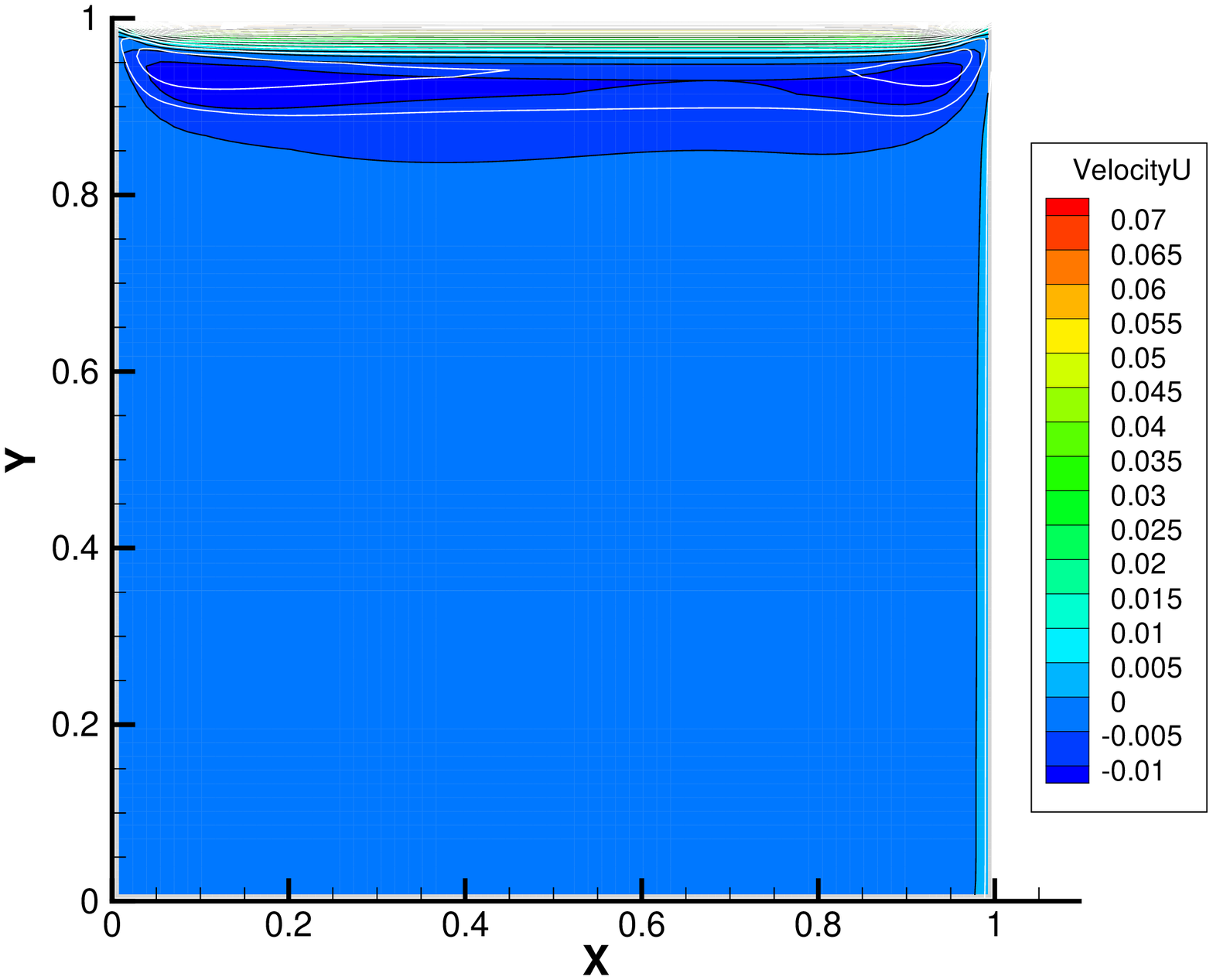}
	\includegraphics[width=0.5\textwidth]{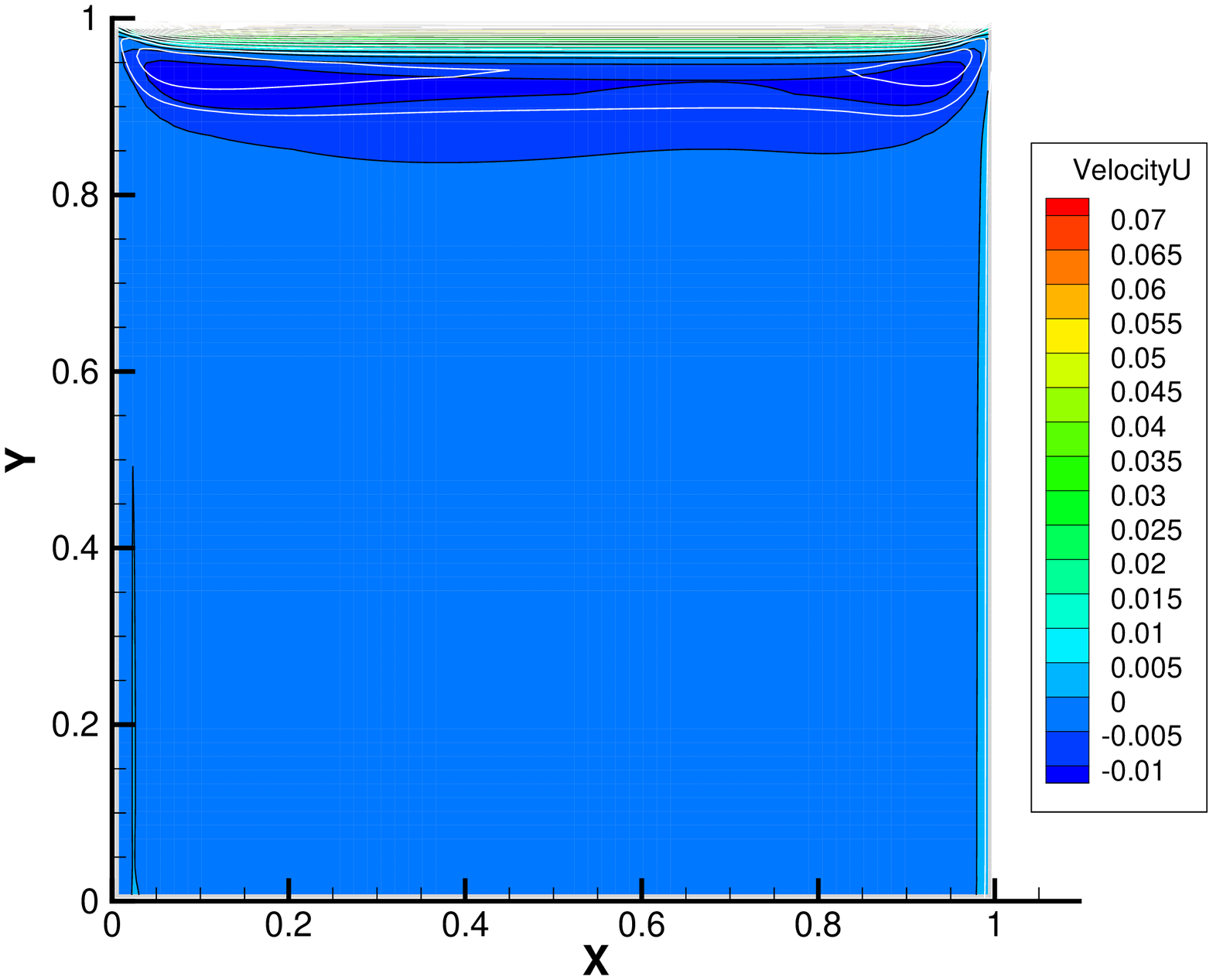}
	\caption{The horizontal velocity distribution of oscillatory cavity flow with Kn = $10^{-3}$ and $St$ = 2. Black lines: original second-order UGKS(left), WENO-AO implemented UGKS(right), white lines: reference result from original second-order UGKS with double mesh number(both).} 
	\label{Fig:osc1e-03}
\end{figure}

\begin{figure}[H]
	\includegraphics[width=0.5\textwidth]{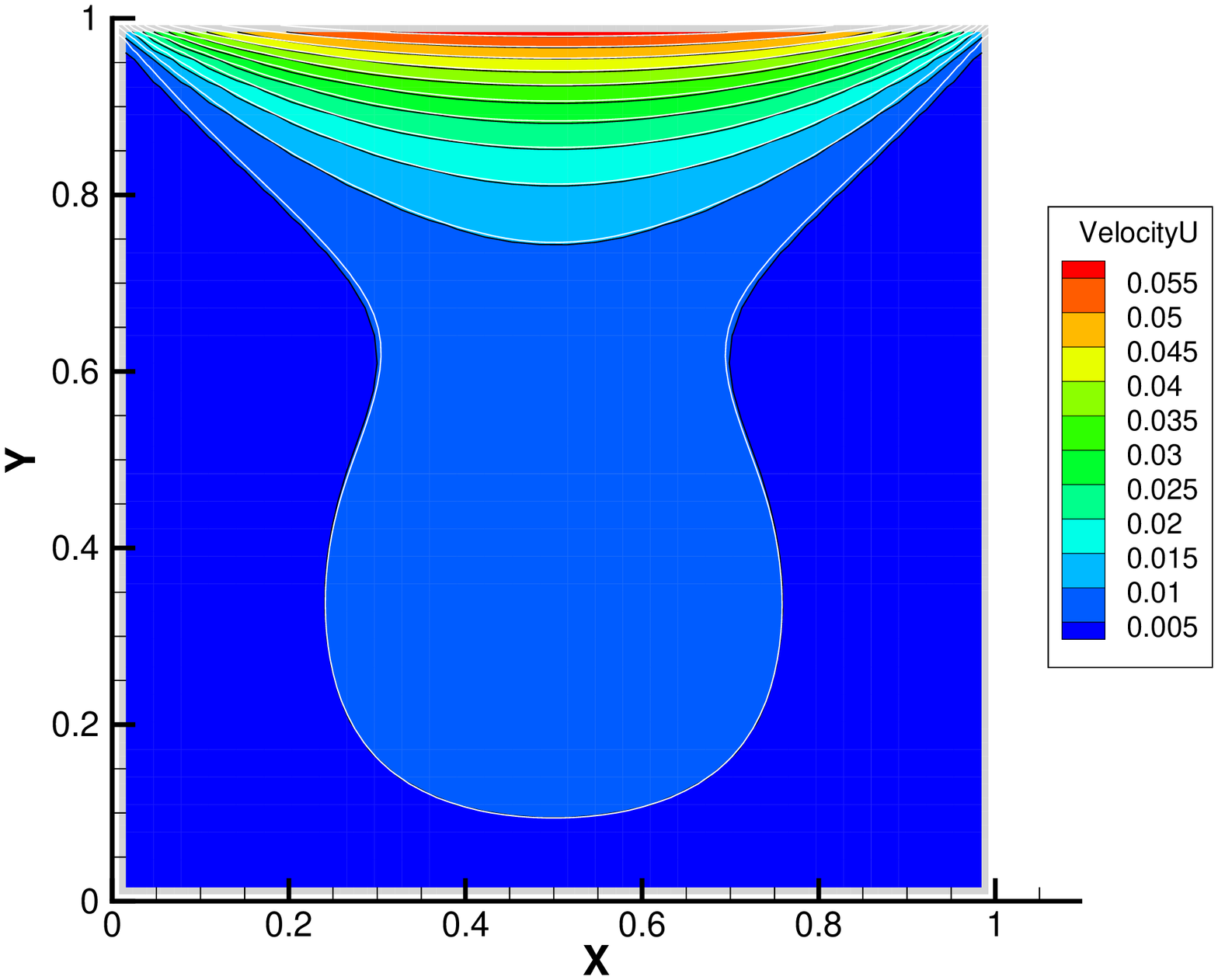}
	\includegraphics[width=0.5\textwidth]{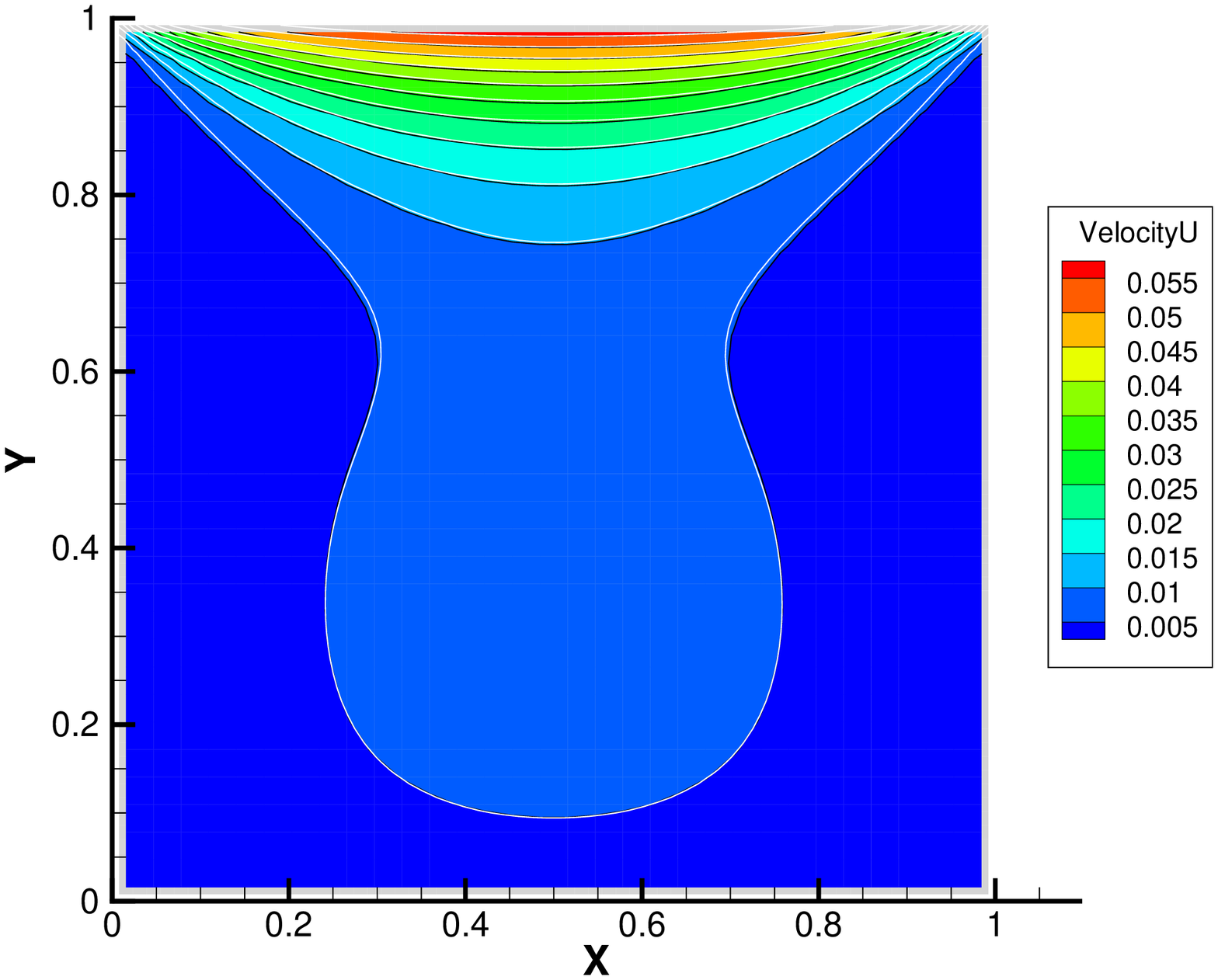}
	\caption{The horizontal velocity distribution of oscillatory cavity flow with Kn = $10^{-1}$ and $St$ = 2. Black lines: original second-order UGKS(left), WENO-AO implemented UGKS(right), white lines: reference result from original second-order UGKS with double mesh number(both).} 
	\label{Fig:osc1e-01}
\end{figure}

For the oscillatory cavity flow, the top plate is replaced from constant velocity lid to the constant frequency oscillatory lid. A non-dimensional parameter, Strouhal number is used for oscillation parameter, which is defined as
\begin{equation}\label{Strouhal_number}
	St=\frac{\omega h}{v_m}
\end{equation}
where $v_m$ = $\sqrt{2RT}$ is the most probable molecular speed. By using Knudsen number and Strouhal number, the rarefaction and oscillation parameter of oscillating cavity flow is controlled. When Strouhal number is high, the frequency of the lid is high. When Strouhal number is low, the frequency of the lid is low. The oscillating cavity tests with different Kn are test at $St$ = 2. The discretization details are applied as same as the previous cavity flow. Due to the periodic characteristic of the test case, all solutions at $t / t_0 = n$, where $n$ is an integer, are evaluated and they are in steady periodic state as same as oscillating Couette flow. 

\begin{figure}[htb!]	
	\includegraphics[width=0.5\textwidth]{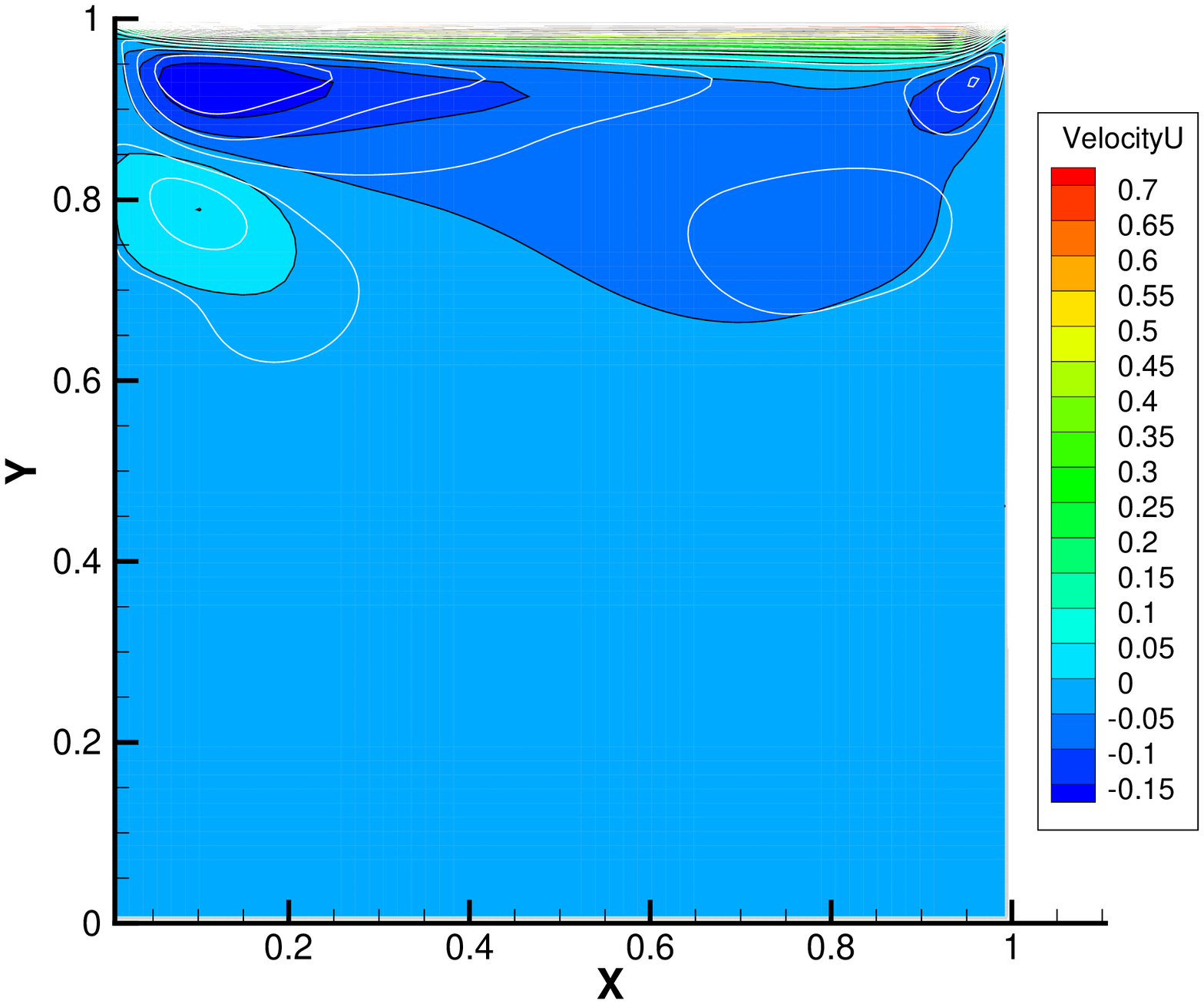}
	\includegraphics[width=0.5\textwidth]{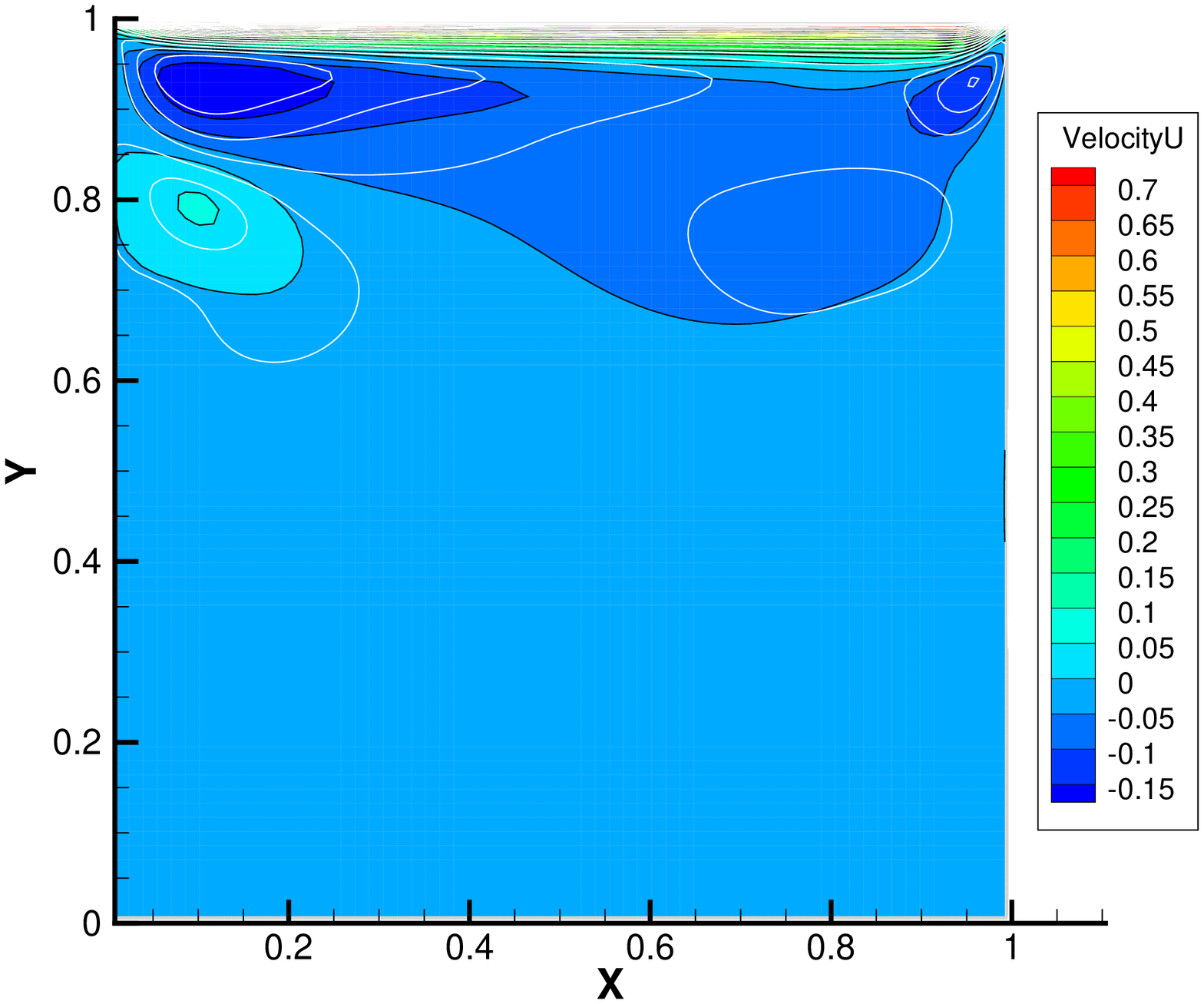}
	\caption{The horizontal velocity distribution of oscillatory cavity flow with Kn = $10^{-3}$ and $St$ = 2 with $U_0$ = 1.0 Mach. Black lines: original second-order UGKS(left), WENO-AO implemented UGKS(right), white lines: reference result from original second-order UGKS with double mesh number(both).} 
	\label{Fig:oscMa1}
\end{figure}

\begin{figure}[htb!]
	\includegraphics[width=0.5\textwidth]{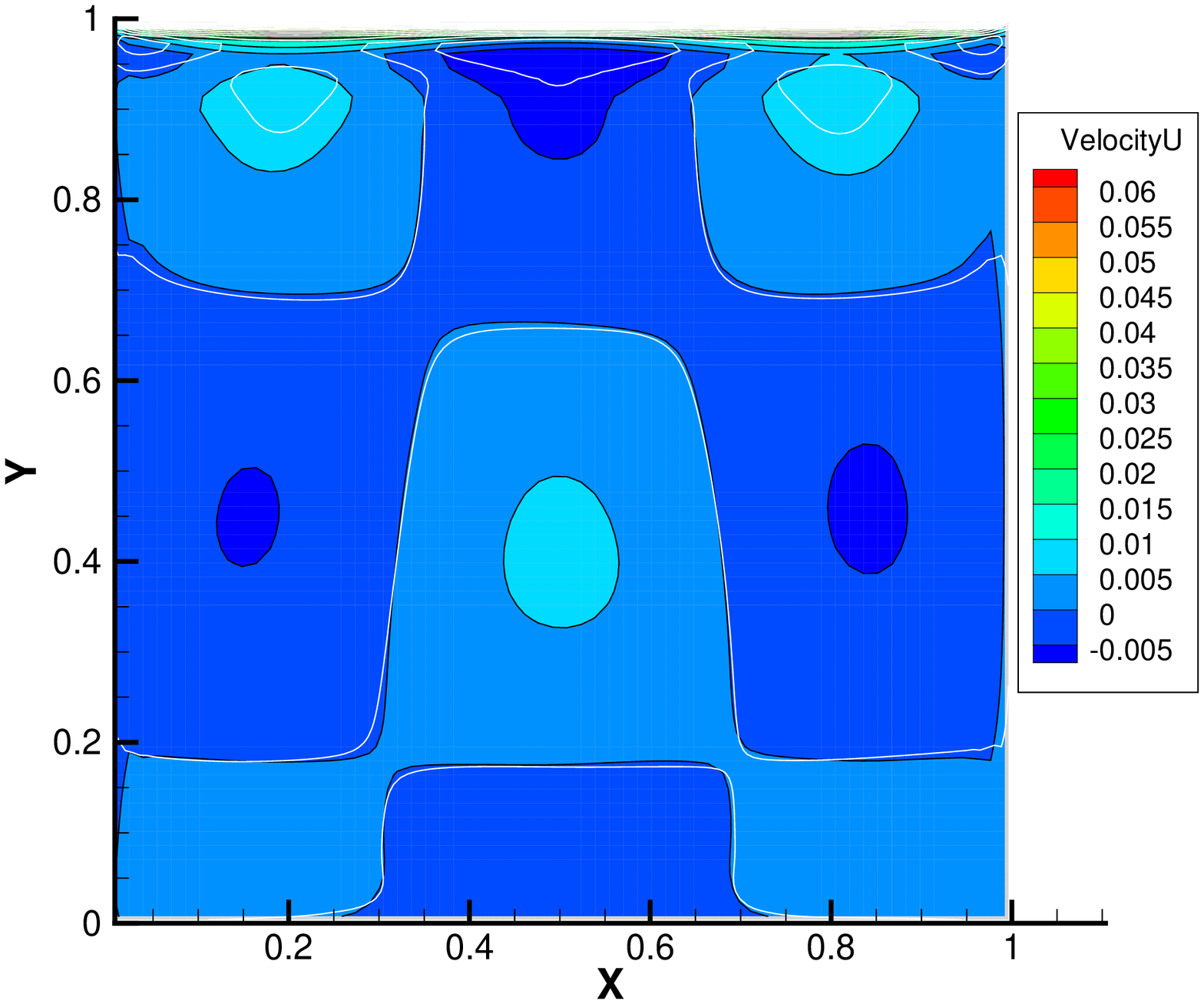}
	\includegraphics[width=0.5\textwidth]{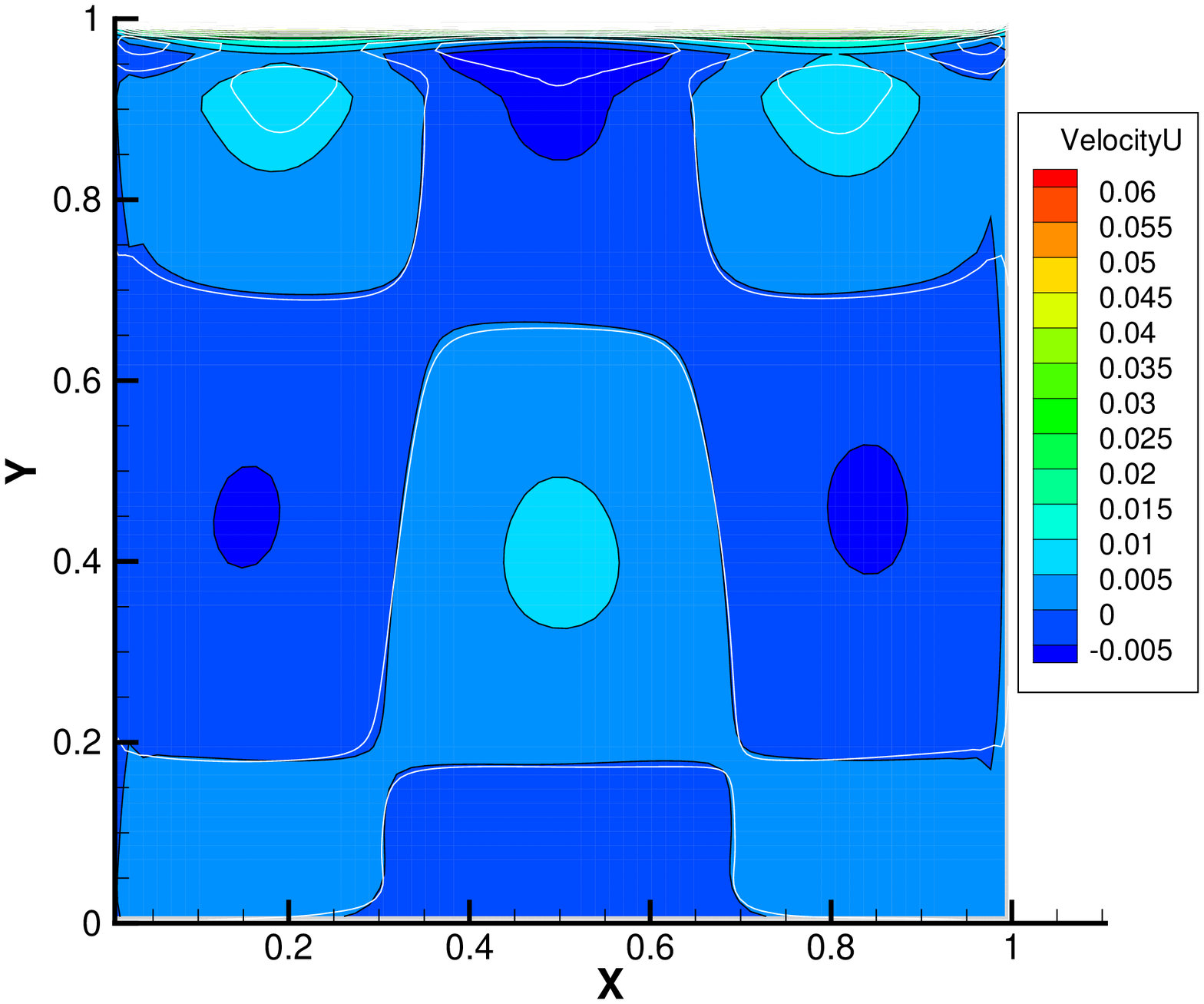}
	\caption{The horizontal velocity distribution of oscillatory cavity flow with Kn = $10^{-3}$ and $St$ = 10 with $U_0$ = 0.1 Mach. Black lines: original second-order UGKS(left), WENO-AO implemented UGKS(right), white lines: reference result from original second-order UGKS with double mesh number(both).} 
	\label{Fig:oscSt10}
\end{figure}

For both Knudsen number cases, according to Figure~\ref{Fig:osc1e-03} and Figure~\ref{Fig:osc1e-01}, the results from WENO-AO implemented UGKS and original second-order UGKS are very close to each other. Since the scheme is targeted to near continuum regime, two more test cases are evaluated at Kn = $10^{-3}$ with different lid velocity and oscillation frequency. Firstly, the cases with the increasing lid velocity from $U_0$ = 0.1 to $U_0$ = 1.0 are evaluated with both original second-order and WENO-AO implemented UGKS in Figure~\ref{Fig:oscMa1}. The WENO-AO implemented UGKS provides better agreement to the reference as expected. The top left part of the contour is described close to the reference with WENO-AO implemented UGKS. Then, the case with increased $St$ = 10.0 are evaluated in Figure~\ref{Fig:oscSt10}. While both results provide close result to each other, it is observed that the velocity contour at the top-left and top-right corner of the WENO-AO implemented UGKS provide slight improvements.

\section{Conclusion}
In this paper, a high-order UGKS is presented for both steady and unsteady solution in all flow regimes. The WENO-AO is applied in the spatial reconstruction of the macroscopic flow variables  for the calculation of equilibrium part of the UGKS, while the discrete distribution function for the non-equilibrium part retains the second-order calculation. With these different treatments of equilibrium and non-equilibrium parts, the increment of computational cost can be well controlled. The two-stage fourth-order method is used for time evolution of the current high-order UGKS.
The current high-order UGKS could recover non-equilibrium flow solutions in rarefied regimes, and obtain better results in the near continuum regimes with higher order accuracy.

The numerical tests of one- and two-dimensional sine wave accuracy test, Sod shock tube test, Couette flow, oscillating Couette flow, cavity flow, and oscillating cavity flow have been computed to validate the current high-order UGKS. The sine wave accuracy test shows that the WENO-AO implemented UGKS can provide higher accuracy. The test cases with different flow regimes proved that the WENO-AO implemented UGKS still maintains the multiscale property of the original UGKS. While it can recover the original second-order UGKS in the highly rarefied flows, it also shows that the scheme can provide better description near the discontinuity and the peak value in the near continuum regime. Furthermore, it is observed that the WENO-AO UGKS can describe the flow better with less number of cells than original second-order UGKS in the near continuum regime due to its higher accuracy in the equilibrium part. In conclusion, the WENO-AO implemented UGKS has potential to give accurate solution and it would be beneficial for calculations in the near continuum regime.

\section{Acknowledgment}
The current research is supported by National Science Foundation of China (11772281, 91852114), and Hong Kong Research Grant Council (16208021). 

\bibliographystyle{plain}
\bibliography{ugks_wenoao}
\end{document}